%% file: main.tex
\documentclass[twocolumn]{aastex62}

\submitjournal{ApJ}

\shorttitle{Tuc III Chemical Analysis}
\shortauthors{Marshall et al.}

\begin{document}
% \vspace*{-\headsep}\vspace*{\headheight}
%  \hfill FERMILAB-PUB-18-651-AE\\
% \vspace*{-\headsep}\vspace*{\headheight}
%  \hfill FERMILAB-PUB-18-651-AE\\
% \vspace*{-\headsep}\vspace*{\headheight}
%  \hfill DES-2017-0289
  
\title{Chemical Abundance Analysis of Tucana III, the Second $r$-process Enhanced Ultra-Faint Dwarf Galaxy\footnote{This paper includes data gathered with the 6.5m Magellan Telescopes located at Las Campanas Observatory, Chile.}}


\input{authors.tex}

%\correspondingauthor{J. L. Marshall}
\email{marshall@tamu.edu}

%\date{\today\\v0}

\begin{abstract}
We present a chemical abundance analysis of four additional confirmed member stars of Tucana III, a Milky Way satellite galaxy candidate in the process of being tidally disrupted as it is accreted by the Galaxy. Two of these stars are centrally located in the core of the galaxy while the other two stars are located in the eastern and western tidal tails.  The four stars have chemical abundance patterns consistent with the one previously studied star in Tucana III: they are moderately enhanced in $r$-process elements, i.e. they have $<$[Eu/Fe]$> \approx +$0.4 dex.  The non-neutron-capture elements generally follow  trends seen in other dwarf galaxies, including a metallicity range of 0.44 dex and the expected trend in $\alpha$-elements, i.e., the lower metallicity stars have higher Ca and Ti abundance. Overall, the chemical abundance patterns of these stars suggest that Tucana III was an ultra-faint dwarf galaxy, and not a globular cluster, before being tidally disturbed.  As is the case for the one other galaxy dominated by $r$-process enhanced stars, Reticulum II, Tucana III's stellar chemical abundances are consistent with pollution from ejecta produced by a binary neutron star merger, although a different $r$-process element or dilution gas mass is required to explain the abundances in these two galaxies if a neutron star merger is the sole source of $r$-process enhancement.
\end{abstract}

\section{Introduction}
\label{section:intro}

Over sixty years ago, \cite{bbfh} summarized a plausible story for the nucleosynthesis of every element in the Periodic Table.  
Since that time, observations of the production processes of all but the heaviest elements have confirmed early theories, with only the production site (or sites) of the rapid neutron-capture, or $r$-process, elements eluding direct observation.  The recent detection of a binary neutron star merger event enabled by LIGO \citep{ligo} and extensive efforts to follow up the event \citep[e.g.][]{drout,shappee} have added a further dimension to the study of $r$-process element production, perhaps enabling the direct observation of the production sites of the heaviest elements for the first time.

In the Milky Way halo and in dwarf galaxies, stars have been found showing large enhancements in $r$-process elements. These are divided into two subclasses: moderately
enhanced $r$-I stars (+0.3 $<$ [Eu/Fe] $<$ +1.0) and highly enhanced $r$-II stars ([Eu/Fe] $>$ +1.0) \citep{bc}. These stars, which are often metal-poor, are quite rare, and as of a few years ago, only $\sim$100 $r$-I and 20 $r$-II stars were known, nearly all located in the halo.
Only recently have $r$-process enhanced stars begun to be found in larger numbers via dedicated searches \citep[e.g.][]{barklem,rpa,sakari2018b} or in serendipitous discoveries during chemical study of Milky Way satellite galaxies \citep{ji_nature, roederer-retII-chem, hansen}.

%Stars enhanced in $r$-process elements are divided into two subclasses: moderately
%enhanced $r$-I stars (+0.3 $<$ [Eu/Fe] $<$ +1.0) and highly enhanced $r$-II stars ([Eu/Fe] $>$ +1.0) \citep{bc}. These metal-poor stars are quite rare, but have been found in the Milky Way halo and in small numbers in dwarf galaxies.  As of a few years ago, only $\sim$100 $r$-I and 20 $r$-II stars were known, nearly all located in the halo.
%Only recently have $r$-process enhanced stars begun to be found in larger numbers via dedicated searches \citep[e.g.][]{barklem,rpa} or in serendipitous discoveries during chemical study of Milky Way satellite galaxies \citep{ji_nature, roederer-retII-chem, hansen}.

Interestingly, two recently discovered Milky Way satellite galaxies, Reticulum II and Tucana III, have been shown to be enhanced in $r$-process elements.  These discoveries have been enabled by modern deep, wide-field imaging surveys such as the Dark Energy Survey \citep[DES;
][]{des}, Magellanic Satellites Survey \citep[MagLiteS;][]{maglites}, Survey of the Magellanic Stellar History \citep[SMASH;][]{smash}, and Pan-STARRS \citep{chambers}, which have revealed faint, previously unknown stellar associations.  
Of particular interest are the discoveries of many dark matter-dominated ultra-faint dwarf galaxies and tidally disrupted stellar streams that have been found in and around the Milky Way halo using images from DES \citep{bechtol_des,koposov_9,adw,kimjerjen_hor2,kim,luque1,luque2,shipp}. 

The first of the DES-discovered ultra-faint dwarf galaxies to be kinematically confirmed as a dark matter-dominated ultra-faint dwarf galaxy was Reticulum II \citep[Ret II;][]{simon_ret2, walker}.  The nine brightest confirmed member stars were subsequently chemically analyzed by \cite{ji_nature,ji_retiicomplete} and 
\cite{roederer-retII-chem}; these authors showed that most of the stars in Ret II are strongly enhanced in the $r$-process elements, i.e., they are $r$-II stars.  Since $r$-II stars are so rare in the Milky Way, it was particularly notable to have found a galaxy seemingly composed primarily of these types of stars.  Even more interesting, the authors conclude that the high fraction of $r$-process enhanced stars must be due to Ret II's chemical history being dominated by a single nucleosynthetic event, most likely a binary neutron star merger. 

A second ultra-faint dwarf galaxy, Tucana III (Tuc III), the subject of this paper, has since been shown to be enhanced in $r$-process elements as well, although to a lower level of enhancement than Ret II.
Tuc III was first identified as a candidate Milky Way satellite galaxy in the DES Year 2 dataset \citep{adw}. 
\cite{simon} measured radial velocities of candidate member stars in Tuc III and used 26 confirmed member stars to show that, if Tuc III is a galaxy, it may be the Milky Way satellite galaxy with the lowest mass and velocity dispersion and also the smallest metallicity dispersion of any known dwarf galaxy \citep[Segue 2 has a similarly low mass][]{seg2}.  Despite the fact that these characteristics place Tuc III 
in a part of parameter space where globular clusters and dwarf galaxies cannot be cleanly separated, \cite{simon} concluded that Tuc III is most likely a dwarf galaxy and not a globular cluster.  A possible explanation for Tuc III's low metallicity and large size is that the prominent central overdensity is actually the center of a previously more populous galaxy that has been tidally stripped, leaving only the core of the galaxy intact.
The brightest confirmed member star in Tuc III has been chemically analyzed by \cite{hansen}, who classified it as an $r$-I star.  

Tuc III is unique amongst recently discovered candidate Milky Way satellites in that the DES discovery images show a linear structure in the filtered stellar density map that extends two degrees to either side of the central overdensity.  The papers reporting the discovery \citep{adw} and kinematic confirmation \citep{simon} of Tuc III suggested that this feature may be consistent with a set of leading and trailing tails resulting from tidal disruption as Tuc III merges with the Milky Way halo.  Indeed, \cite{ting} have recently confirmed that these structures are kinematically associated with the Tuc III system, adding 22 confirmed members of the Tuc III tidal tails to the 26 central core stars confirmed by \cite{simon}.  The tidal tails extend at least 2 degrees to either side of the core and show a significant velocity gradient across the structure, as expected for a system being tidally disrupted as it merges with the Milky Way.  Furthermore, \cite{erkal} used these same stars and their measured velocity gradient along with predicted space velocities to fit an orbit about the Milky Way, indicating that Tuc III has had a recent close passage with the Large Magellanic Cloud.  This prediction was further refined with Gaia proper motions \citep{simon2018} which demonstrate that Tuc III is now on a highly eccentric orbit around the Milky Way with a pericenter of $\sim$3 kpc.

In this paper we present the chemical abundance analysis of four additional stars in the Tuc III stellar system; two located in the core of the galaxy and two in the tidal tails.  
This paper is organized as follows: In Section~\ref{section:obs} we describe the observations of the four stars.  We describe the radial velocity and abundance measurements of these stars in Section \ref{section:abund} and present the results of these measurements in Section \ref{section:results}. In Section \ref{section:discuss} we discuss the implications of the chemical abundance patterns of the Tuc III member stars and in Section \ref{section:concl} we conclude.

\section{Observations and Data Reduction}
\label{section:obs}
We selected a sample of four confirmed member stars of Tuc III that had not been studied previously with high resolution spectroscopy: two stars were selected from the sample of confirmed member stars of \cite{simon}, located in the core of the galaxy, and two stars were selected in Tuc III's tidal tails from the sample of confirmed member stars of \cite{ting}. 
Throughout this work we include one star previously chemically analyzed by \cite{hansen} for reference.

\begin{figure}[hbt!]
\centering 
\includegraphics[scale=0.6]{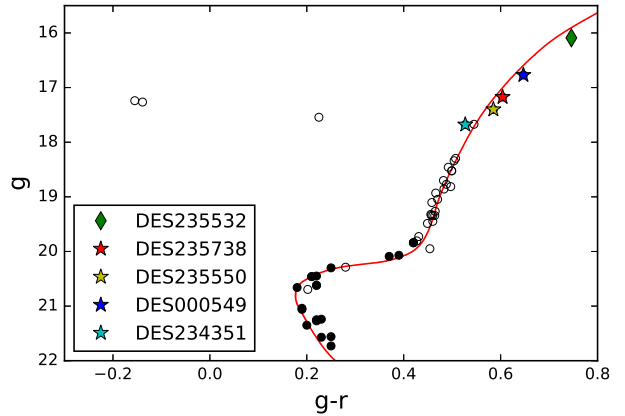}
\caption{
Color-magnitude diagram composed with DES photometry for confirmed member stars of Tuc III.  DES J235532 was shown to be an $r$-I star by \cite{hansen} and is marked with a green diamond; the four stars studied in this work are marked with stars: the red and yellow stars are located in the core of the galaxy; the blue and magenta stars are in the tails.  Filled circles mark other confirmed member stars in the core of Tuc III from \cite{simon}; open circles are confirmed member stars located in the core and tidal tails from \cite{ting}.  A Dartmouth isochrone \citep{dotter} of a stellar population having an age of 12.5 Gyr, [Fe/H]=$-2.3$, [$\alpha$/Fe]=$0.2$ and a distance of 25 kpc is overplotted (red curve). 
}
\label{fig:cmd}
\end{figure}

\begin{figure}[hbt!]
\centering 
\includegraphics[scale=0.6]{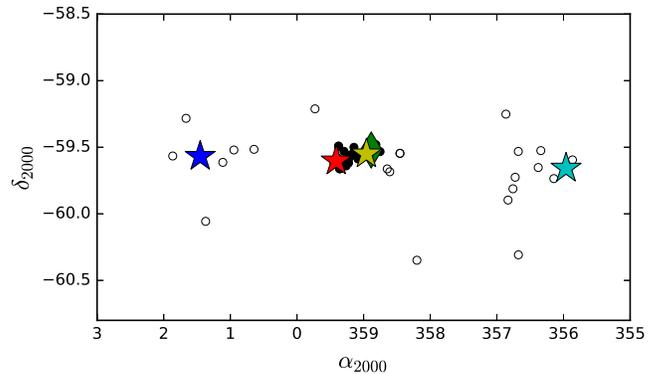}
\caption{
Angular distribution of confirmed member stars in the Tuc III core and tidal tails. Symbols as in Figure~\ref{fig:cmd}.
}
\label{fig:map}
\end{figure}

A color-magnitude diagram of the Tuc III member stars is presented in Figure~\ref{fig:cmd}, with reddening-corrected stellar magnitudes from \cite{ting} and \cite{simon}. Figure~\ref{fig:map} shows the locations of these five stars  with respect to confirmed member stars in Tuc III's core \citep{simon} and tail \citep{ting}. In Figures~\ref{fig:cmd} and \ref{fig:map}, astrometry and photometry are those reported by \cite{ting} when available, since that more recent work presents measurements from the DES DR1 public data release \citep{dr1}, an updated, better-calibrated version of the DES catalog than the Y2Q1 catalog \citep{adw} used by \cite{simon}. %Note however that the brightest stars are not included in DES DR1 since they are saturated in the coadded images and the photometry was calculated for those stars from DES single epoch images.

Observations were performed with the MIKE spectrograph \citep{bernstein-MIKE} at the Magellan-Clay Telescope at Las Campanas Observatory.  Observations took place on 05--07 August 2016. We used a 0.7 arcsec slit with 2x2 pixel binning to obtain a spectral resolution of R=$\lambda / \Delta\lambda\approx$ 41,000 in the blue and 32,000 in the red. The spectra cover 3310\AA$<\lambda<$5000\AA\ in the blue channel and 4830\AA$<\lambda<$9160\AA\ in the red. 

Conditions were somewhat marginal on the first two nights, with some cloud cover and seeing of 0.8 to 1.2 arcsec; clouds cleared and seeing improved to 0.7 arcsec on the third night.  Each star was observed on only one night, with multiple 30-minute integrations interspersed with ThAr comparison lamp spectra at intervals of no more than one hour to facilitate precise wavelength calibration and radial velocity measurements.  In addition to the program stars, at least one radial velocity standard star was observed on each night; telluric standards were observed on the first and third nights.

An observing log is given in Table~\ref{obslog}. The four stars studied here will be referred to as DES J235738, DES J235550, DES J000549, and DES J234351 for brevity. We include the star DES J235532 studied by \cite{hansen} for reference. Also included in Table~\ref{obslog} are the DES astrometry and $g$ and $g-r$ photometry for each star reported by  \cite{ting}. 

Reduction of the data, including bias subtraction, flat fielding, spectral extraction, wavelength calibration, and coadding was completed on the mountain with the latest version of the MIKE pipeline \citep{kelson-MIKE-pipeline}. 
Formal signal-to-noise ($S/N$) ratios were measured at 4100 and 5500\AA\ using the IRAF task \textit{splot} and are presented in Table~\ref{obslog}.

\begin{deluxetable*}{lccccccccc}[htb]
\tablecaption{Observing log\label{obslog}} 
\tablehead{Object Name & R.A.\tablenotemark{a} & Dec.\tablenotemark{a} & $g$\tablenotemark{a} & $g-r$\tablenotemark{a} & Date Observed\tablenotemark{b} & $t_{exp}$ & $S/N$ &$S/N$& Location \\
&(J2000)&(J2000)&(mag)&(mag)&(MJD)&  (hr)&at 4200\AA\ &at 5500\AA\ &}
\startdata
DES J235532-593115\tablenotemark{c} &23:55:32.7  & $-$59:31:15.0 & $16.090$ & 0.746 &57248 & 3.5&20&50& Core  \\
\hline
DES J235738-593612 & 23:57:38.5 & $-$59:36:11.7 & $17.173$ & $0.604$&57605.69 &2 &20 &30& Core  \\
DES J235550-593300 & 23:55:49.9 & $-$59:33:00.0 & $17.400$ & $0.585$ &57607.77 & 2 &20&30& Core  \\
DES J000549-593406 & 00:05:48.7 & $-$59:34:06.1 & $16.770$ & $0.647$ &57605.78 & 4.5 &30&40& Tail   \\
DES J234351-593926 & 23:43:50.8 & $-$59:39:25.6 & $17.678$ & $0.527$ &57606.82 & 2.5 &10&20& Tail   \\
\enddata
\tablenotetext{a}{Astrometry and dereddened photometry from \cite{ting}.}
\tablenotetext{b}{Reported at the midpoint of the observation.}
\tablenotetext{c}{From \cite{hansen}, included for reference.}
\end{deluxetable*}

\section{Stellar Parameter Determination and Chemical Abundance Analysis}
\label{section:abund}

\subsection{Radial Velocities}

Radial velocities for each star were measured by comparing the program star with a radial velocity standard star (HD136202) observed on the first night of the run.  Radial velocities were derived via cross-correlation of each order of the program star spectrum with the corresponding order of the standard star spectrum.  The blue and the red arms of the spectrograph were considered independently.  The mean values of the resulting relative velocities from each arm, for 28 orders in the blue and 6 orders in the red, with 3-$\sigma$ outliers rejected, e.g. $RV_{blue}=\sum{(v_{n, blue}) / N}$, where $v_{n, blue}$ is the velocity derived from each order of the blue arm and $N$ is the total number of blue orders.   These velocities were then averaged to form the final reported velocity for each star: $RV_{final}=(RV_{blue} + RV_{red})/2$.  Errors were derived using the standard deviation of the individual velocities determined from each order, again considering the blue and red arms separately.  The reported error on each velocity is calculated as $\sigma_{RV_{final}} = (\sigma_{RV_{blue}}^2/2 + \sigma_{RV_{red}}^2/2)^{1/2}$.  Measured radial velocities for all four stars are presented in Table \ref{params}. The radial velocity measurements were used to place each program star spectrum on a wavelength scale associated with rest wavelengths.  These velocities can also be used to investigate whether the stars are in fact binaries, as discussed in Section \ref{sec:binary}.

\begin{deluxetable*}{lccccc}[!htb]
\caption{Measured Stellar Parameters\label{params}
}
\tablehead{ID &v$_{hel}$& $T_{eff}$ & log $g$ & v$_{micro}$  & [Fe/H] \\
&(km s$^{-1}$)&(K)&&(km s$^{-1}$)&
}
\startdata
DES J235532 &$-103.4 \pm0.3$ & $4720 \pm 100$ & $1.33 \pm 0.3$ &$2.0\pm0.3$& $-2.25 \pm 0.18$ \\
\hline
DES J235738 &$-100.9\pm0.8$& $4720 \pm 150$ & $1.36\pm 0.3$ & $1.3\pm 0.3$ & $-2.58\pm0.18$ \\
DES J235550 &$-102.9\pm0.8$& $4720 \pm 150$ & $1.55\pm 0.3$ & $1.5\pm 0.3$ & $-2.69\pm0.17$ \\
DES J000549 &$-92.6\pm0.9$& $4675 \pm 150$ & $1.39\pm 0.3$ & $1.7\pm 0.3$ & $-2.61\pm0.08$ \\
DES J234351 &$-121.5\pm1.1$& $4900 \pm 150$ & $1.88\pm 0.3$ & $1.6\pm 0.3$ & $-2.69\pm0.14$ \\
\enddata
\end{deluxetable*}

\subsection{Stellar parameter measurements and abundance analysis}
Stellar parameters for the four stars were derived spectroscopically, following the method described by \citet{hansen}.  In brief, we used the 2017 version of the MOOG spectral synthesis program \citep{moog-sneden}, making the assumption of local thermodynamic equilibrium and including Rayleigh scattering treatment as described by \citet{sobeck2011}\footnote{https://github.com/alexji/moog17scat}. Initial effective temperatures were determined from excitation equilibrium of \ion{Fe}{1} lines and thereafter placed on a photometric scale using the relation from \cite{frebel2013}.   Following this the surface gravities ($\log g$) were determined from ionization equilibrium between the \ion{Fe}{1} and \ion{Fe}{2} lines. Finally microturbulent velocities ($\xi$) were determined by removing any trend in line abundances with reduced equivalent widths for the \ion{Fe}{1} lines. For the four stars studied here, J235738, J235550, J000549, and J234351, we were able to use 103, 122, 157, and 90 \ion{Fe}{1} lines and 15, 13, 20, and 16 \ion{Fe}{2} lines, respectively, for this analysis.  Final stellar parameters are presented in Table \ref{params} and lines used for the analysis of each star are listed in Table \ref{tab:ew}.

\input{Fe.tab}

Abundances were derived from equivalent width measurements and spectral synthesis using MOOG. We used  $\alpha$-enhanced ($\mathrm{[\alpha/Fe]} = +0.4$) 1D LTE ATLAS9 model atmospheres \citep{castelli2003} and the solar photosphere abundances from \cite{asplund-solar}.  Line lists were generated using the linemake package\footnote{https://github.com/vmplacco/linemake} (C.\ Sneden, private comm.), including molecular lines for CH, C$_2$, and CN and isotopic shift and hyperfine structure information. Measured stellar abundances are presented in Table \ref{abund}.

We note here that the effective temperature derived for three of these stars (including J235532) are all equal, which is unexpected given that J235532 is somewhat redder than the other two stars.  We have carefully reconsidered the derived effective temperatures for all five stars, including J235532, and find no errors in the analysis. Expected values from the Dartmouth isochrone (shown in Figure~\ref{fig:cmd}) suggest a temperature difference of $\Delta T \sim$300 K between stars having the colors of J235532 and J235550, which agrees with our derived spectroscopic temperatures within errors.

Uncertainties on the derived abundance for J000549 arising from stellar parameter uncertainties are listed in Table \ref{error}. These uncertainties were computed by deriving abundances with different atmospheric models, each with one parameter varied by its uncertainty as given in Table \ref{params} and added in quadrature. As all stars have similar stellar parameters and spectral quality we consider these uncertainties to be applicable to all four stars studied here.

\begin{figure*}[htb!]
\centering 
\includegraphics[scale=0.48]{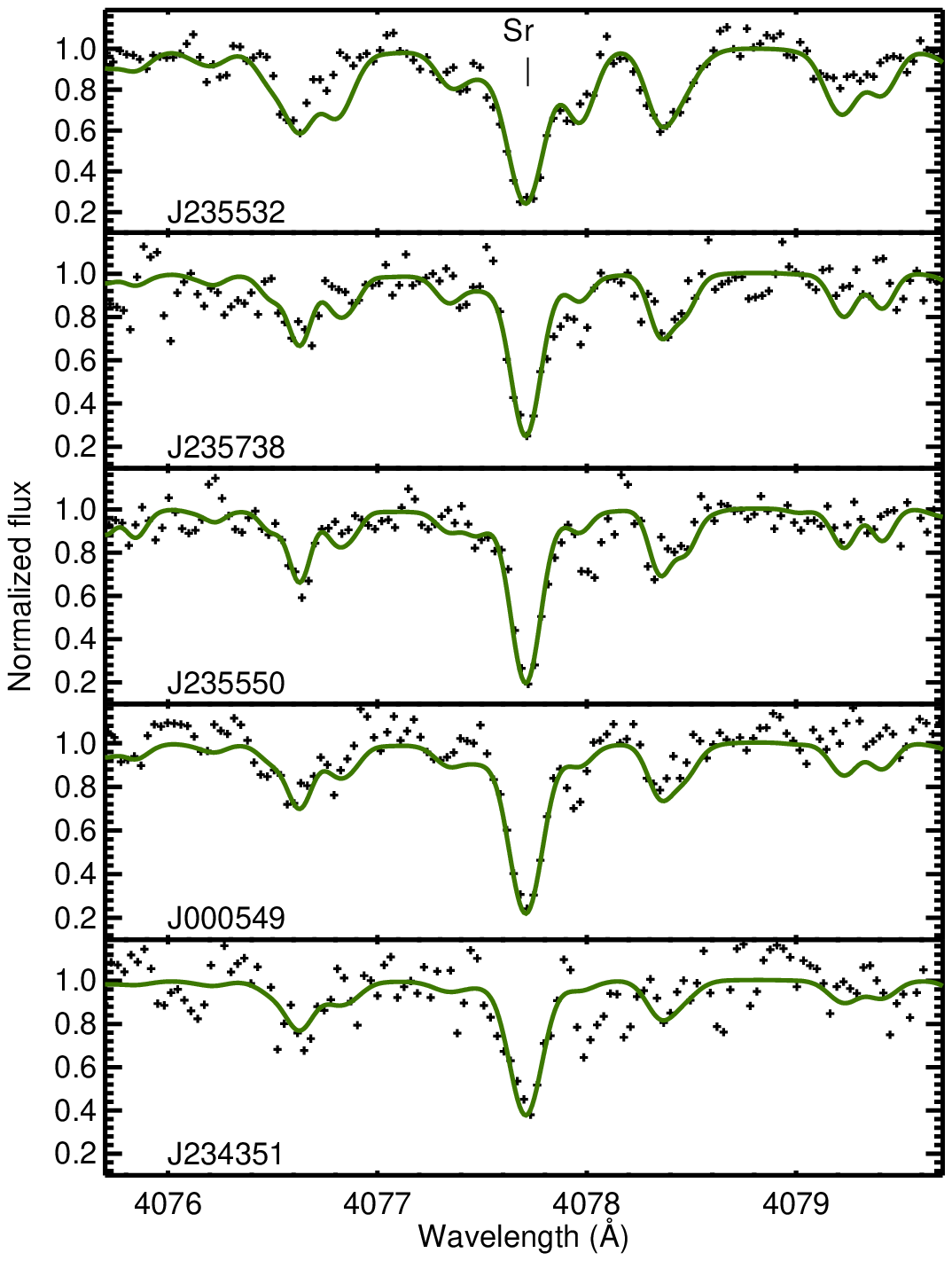}
\includegraphics[scale=0.48]{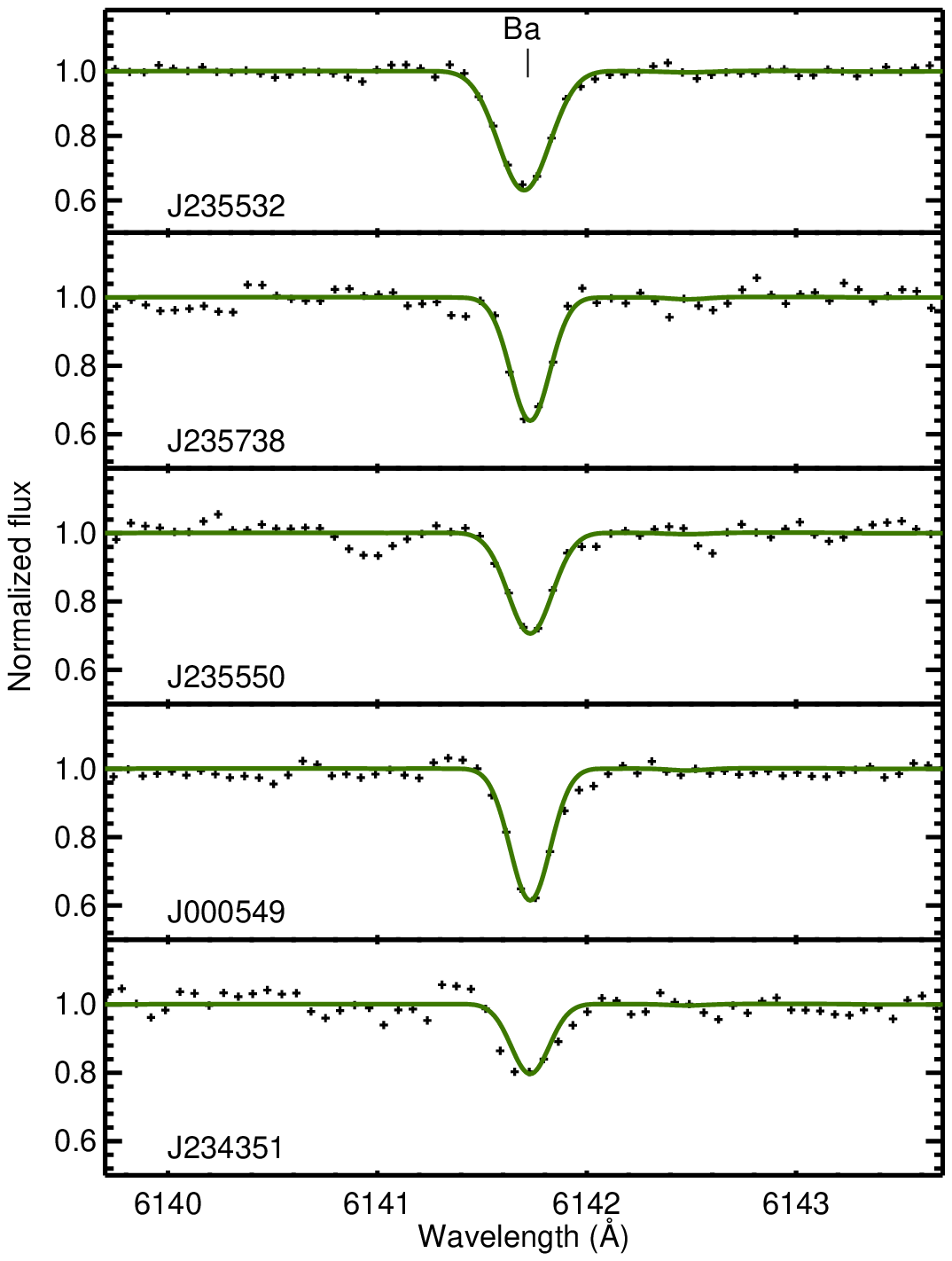}
\includegraphics[scale=0.48]{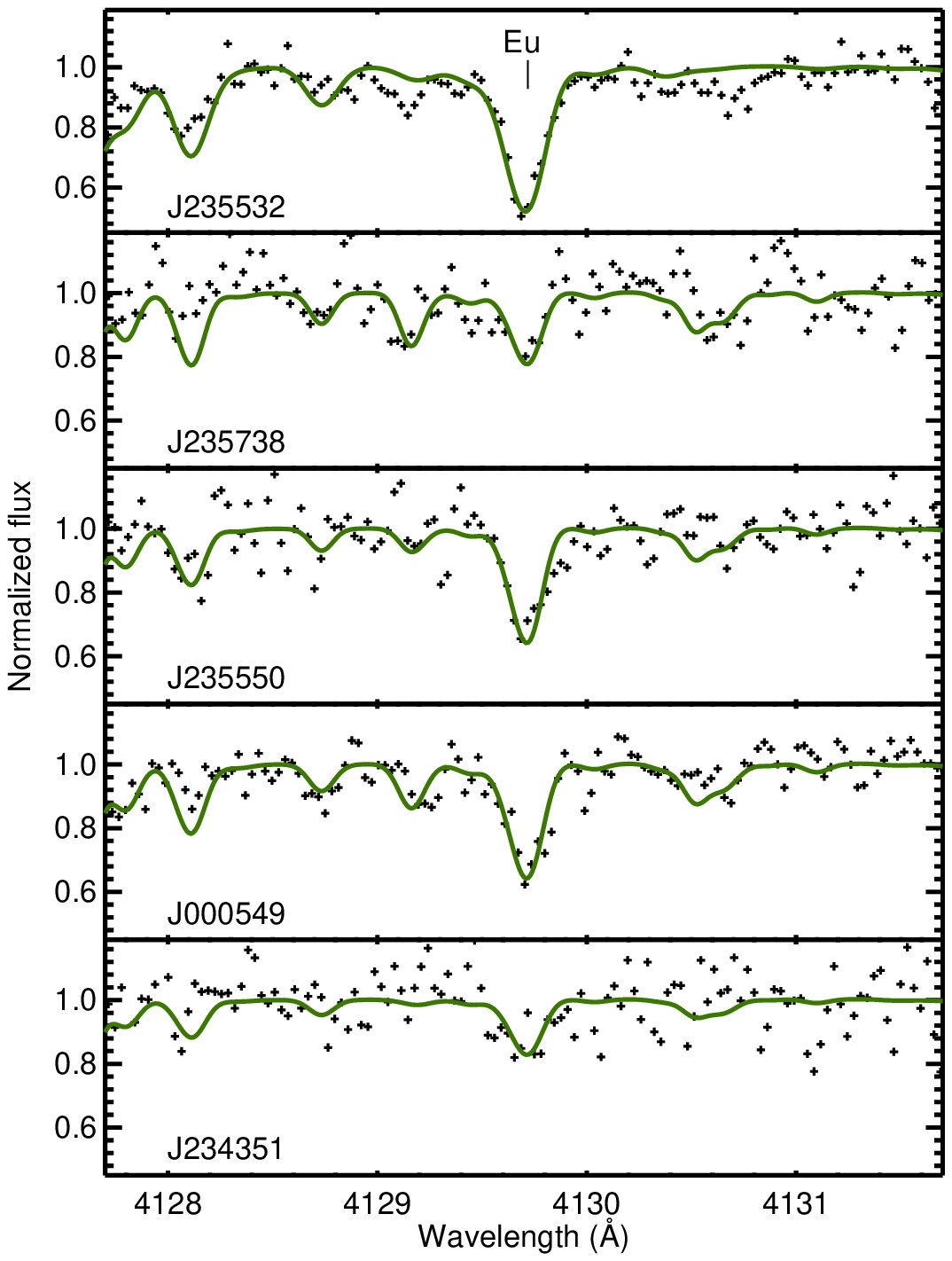}
\caption{Synthesis of the neutron-capture elements Sr, Ba, and Eu in five Tuc III stars.  Top panel: J235532 reproduced from \cite{hansen} for reference; lower four panels: J235738 (top), J235550 (middle top), J000549 (middle bottom), and J234351 (bottom), this work.
}
\label{fig:synth}
\end{figure*}

Sample synthetic spectra for absorption lines of Sr, Ba and Eu can be found in Figure \ref{fig:synth}, overlaid onto the observed spectra.

\section{Results}\label{section:results}

%\subsection{$\alpha$ and iron peak elements}
\subsection{Non-neutron-capture elements}
\begin{figure*}[htb!]
\centering 
\includegraphics[scale=0.5]{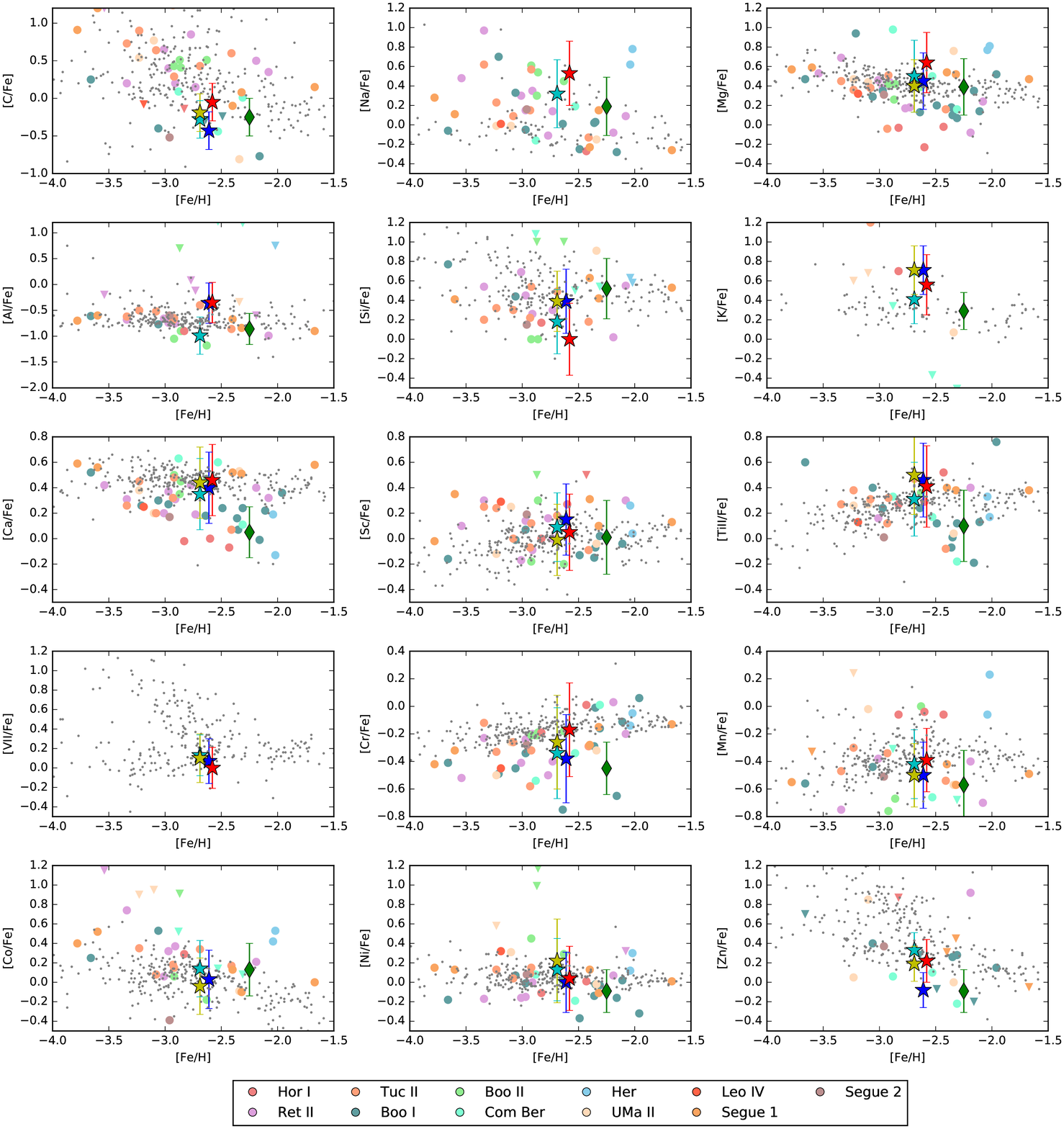}
\caption{
Non-neutron-capture element abundances in Tuc III (symbols as in Figure~\ref{fig:cmd}) compared to stars in other ultra-faint dwarf galaxies (colored circles and triangles, the latter indicate upper limits): Horologium I \citep{nagasawa}, Reticulum II \citep{ji_retiicomplete}, Tucana II \citep{tucii}, Bootes I \citep{gilmore_booi,ishigaki_booi,frebel_booi,norris_booi}, Bootes II \citep{booii},Coma Berenices and Ursa Major II \citep{comber_umaii}, Hercules \citep{her}, Leo IV \citep{leoiv}, Segue 1 \citep{segue1}, and Segue 2 \citep{roedererkirbysegue2}. Milky Way halo stars \citep{roed} are plotted as small grey points. The expected trend in $\alpha$-elements can be seen most easily in the Ca and Ti abundances.  
}
\label{fig:all}
\end{figure*}

We derive abundances for eighteen non-neutron-capture elements from C to Zn, including a range of $\alpha$-elements (Mg, Si, Ca, Ti) and iron peak (Cr, Mn, Fe, Co, Ni) elements. Figure \ref{fig:all} shows the abundances of fifteen of these elements as a function of [Fe/H] compared to stars in other ultra-faint dwarf galaxies and stars in the Milky Way halo.
%Abundance trends in the non-neutron-capture elements is as expected and compares well with stars in other metal-poor populations.  Specifically, with 
When considering the sample of all five stars, a range of metallicity ([Fe/H]) is observed as well as the expected trend in $\alpha$-elements, i.e. that lower metallicity stars in the stellar population have higher $\alpha$ abundances since they were formed at a time at which stellar nucleosynthesis was dominated by Type II SNe, compared to the more metal-rich stars that are formed later after the stellar population has been polluted by Type Ia SN explosions \citep{tinsley}. This  $\alpha$ ``knee'' is observed in the elements Ca and Ti; see Figure \ref{fig:all}.

None of the stars can be classified as carbon-enhanced metal-poor stars (CEMP; $\mathrm{[C/Fe]} > 0.7$).

\subsection{Neutron-capture elements}
\begin{figure*}[htb!]
\centering 
\includegraphics[scale=0.5]{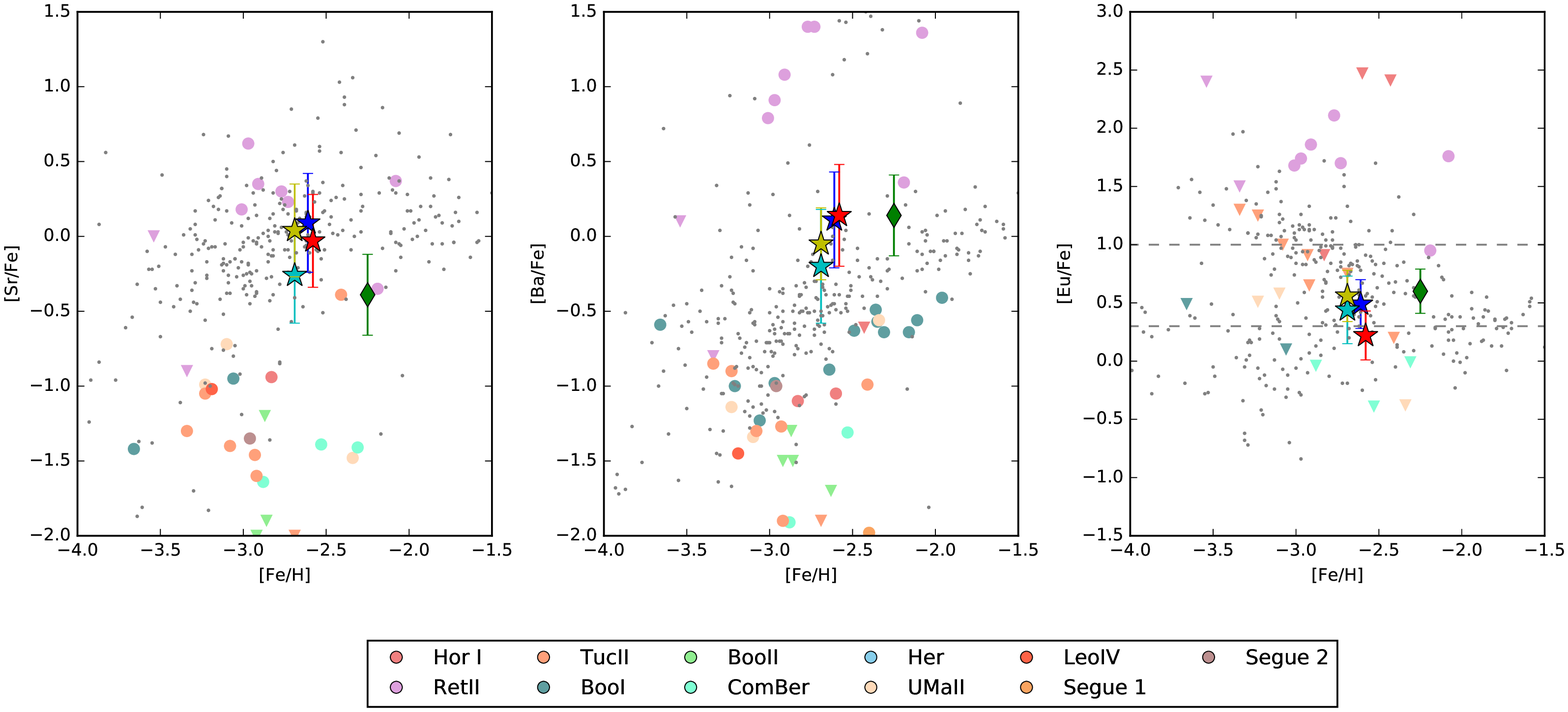}
\caption{
Neutron-capture element abundances in Tuc III compared to other ultra-faint dwarf galaxies and Milky Way halo stars. Symbols as in Figure \ref{fig:all}. Dashed lines in the right panel indicate the traditional definition of $r$-process enhanced stars: $r$-I stars have 0.3 $<$ [Eu/Fe] $<$ 1; $r$-II stars have [Eu/Fe] $>$ 1. Four of the five Tuc III stars lie within these boundaries and are classified as $r$-I stars; the fifth star (J235738) has error bars that cross the discriminator.
}
\label{fig:neutron}
\end{figure*}

We derive abundances or upper limits for eleven neutron-capture elements from Sr to Er. Figure \ref{fig:neutron} shows abundances of the neutron-capture elements Sr, Ba, and Eu compared to stars in other ultra-faint dwarf galaxies and stars in the Milky Way halo. 

\begin{figure*}[htb!]
\centering 
\includegraphics[scale=0.6]{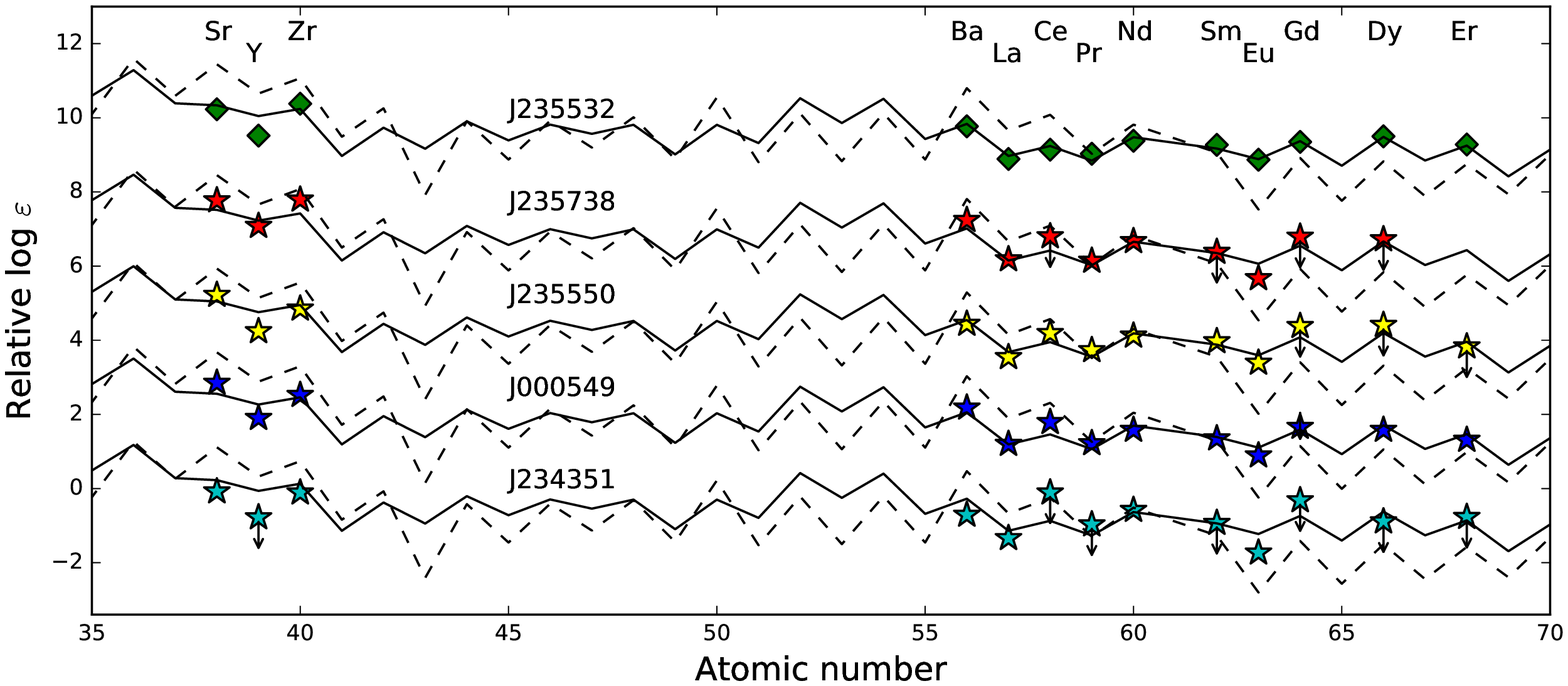}
\caption{
Absolute abundances of neutron-capture elements for our stars compared to scaled solar system $s$-process (dashed line) and $r$-process (solid line) abundance patterns, taken from \cite{simmerer}. A constant offset has been added to each star's abundances.}
\label{fig:rncap}
\end{figure*}

Figure \ref{fig:rncap} shows the neutron-capture element abundance pattern for each of the stars compared to the solar system $s$- and $r$-process abundance pattern from \cite{simmerer}. For each star the solar pattern has been scaled to the average residual between the star and the Sun for elements with abundances (not upper limits) from Ba to Er. It is clear from Figure \ref{fig:rncap} that the neutron-capture abundance pattern in four of the stars is better matched by the solar system $r$-process abundance pattern than the $s$-process pattern. J234351, with abundances measured for only six neutron-capture elements, is the exception. Neither the solar system $r$- nor $s$-process pattern matches the derived abundances for this star well. Note however that this star's spectrum has the lowest signal-to-noise of those studied here.
%This could be explained if this star is confirmed to be a binary, as we discuss in Section \ref{sec:binary}; note however that this star's spectrum also has the lowest signal-to-noise of those studied here, so additional data are required. 

We adopt the \cite{hansen} definition of $r$-process enhanced stars: $r$-I stars are defined to have 0.3 $<$ [Eu/Fe] $<$ 1 and [Eu/Ba]$> $0.4, while $r$-II stars have [Eu/Fe] $>$ 1 and [Eu/Ba] $>$ 0.4, with the additional constraint on [Eu/Ba] added to the traditional definition of $r$-process enhancement in order to ensure that the enhancement is entirely due to elemental production via the $r$-process, and is not confused by contributions from the $s$-process. Since Eu is nearly entirely produced in the $r$-process \citep[94\%,][]{kochedv} but Ba is produced primarily in the $s$-process \citep[85\% according to][]{burris}, the ratio of Eu/Ba is often used to gauge whether the neutron-capture elements in a given star were produced primarily via the $s$- or $r$-process.

\begin{figure}[!htb]
\centering 
\includegraphics[scale=0.6]{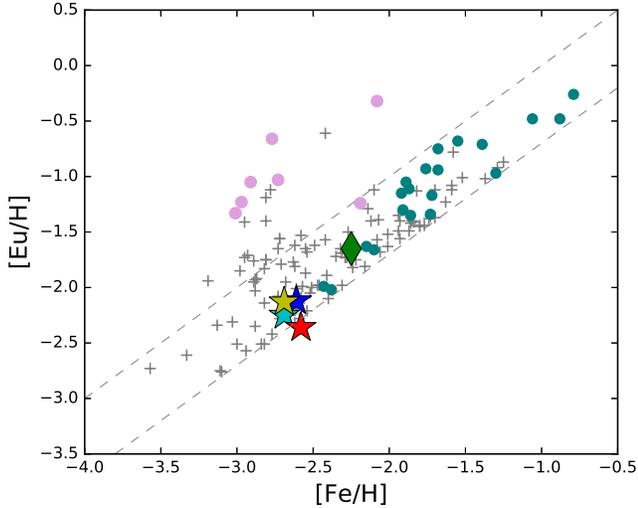}
\caption{
[Eu/H] as a function of [Fe/H] for the stars in Tuc III (symbols as in previous figures) compared to $r$-process enhanced stars in the halo (grey plusses) and other dwarf galaxies, including classical dwarfs \citep[green dots; references given in][]{hansen} and Reticulum II (plum circles). Dashed lines show limits for $r$-I and $r$-II stars. The majority of stars in the halo and in other galaxies are not $r$-process enhanced; these would appear in the lower right and are not plotted here.
}
\label{fig:EuH}
\end{figure}

Our measurements show that overall Tuc III is moderately enhanced in $r$-process elements: three of the four stars studied here are $r$-I stars having 0.3 $<$ [Eu/Fe] $<$ 1, as is J235532, the Tuc III star previously studied by \cite{hansen}.
Figure \ref{fig:EuH} shows the $\mathrm{[Eu/H]}$ ratios as function of metallicity for the five Tuc III stars along with $r$-process enhanced stars in other dwarf galaxies and in the Milky Way halo. 

\subsection{Velocity gradient}

\begin{figure}[!htb]
\centering 
\includegraphics[scale=0.6]{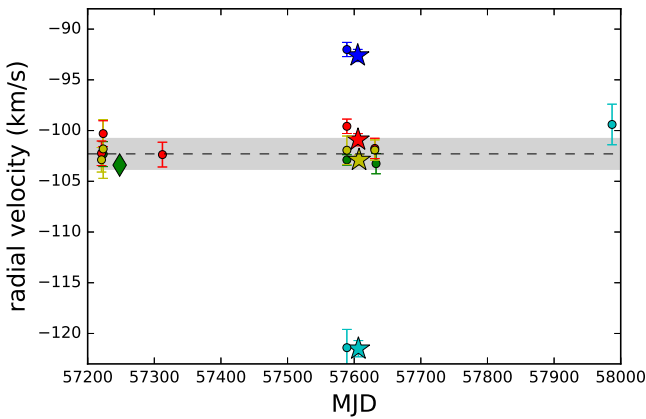}
\caption{
Radial velocities of the five stars as a function of time. Velocities measured by IMACS \citep{simon} and AAT \citep{ting} are marked as circles; symbols for the stars considered in this work are as in Figure~\ref{fig:cmd}. The dashed line and shaded region show the mean radial velocity and upper limit on the velocity dispersion of the Tuc III core stars as measured by \cite{simon}. The radial velocities of the tail stars confirm the velocity gradient measured in the AAT data. DES J234351 shows significant velocity variation and is likely in a binary system; DES J235738 may also be a binary.
}
\label{fig:rv}
\end{figure}

Measured radial velocities of the stars studied here are consistent with \cite{simon} and \cite{ting} as shown in Figure \ref{fig:rv}.  We collect all measured radial velocities for the five Tuc III stars in Table~\ref{rvtable}.  We confirm the results of \cite{ting}, i.e. that there is a significant velocity gradient across the Tuc III system.  We also see some evidence for velocity variations in individual stars, discussed in more detail in Section \ref{sec:binary}.

\section{Discussion}
\label{section:discuss}

Throughout this Section we include the star DES J235532 studied by \cite{hansen} in the discussion, increasing the sample size to five.

\subsection{The $r$-process enhancement event}

Tuc III is the second ultra-faint dwarf galaxy containing multiple  stars enhanced in $r$-process elements. As discussed in Section \ref{section:intro}, the first such galaxy, Ret II, is even more highly enhanced than Tuc III, i.e. many of the Ret II stars are $r$-II stars (see Figure \ref{fig:EuH}).  The fact that so many stars in a galaxy as small as Ret II share a common chemical pattern suggests that a single nucleosynthetic event must have occurred early in the history of the galaxy, polluting future generations of stars, and that most of the stars in the galaxy were impacted by the event.
 
The $r$-II stars in Ret~II have an average enhancement in Eu of $\mathrm{[Eu/H]} \sim -1$. \cite{ji_nature} used this level and an estimated dilution gas mass, i.e.~the mass of hydrogen gas that the $r$-process material is diluted into, of Ret~II of $10^6$ M$_{\odot}$ to argue that a binary neutron star merger was the most likely source of the excess of Eu detected in Ret~II as other sources would not have produced enough Eu to enhance the galaxy to the level detected. In Tuc~III we find an average enhancement in Eu of $\mathrm{[Eu/H]} \sim -2$. Assuming the same binary neutron star merger Eu ejecta mass as in \cite{ji_nature}, $10^{-4.5}$ M$_{\odot}$, leads to a dilution gas mass of $\sim 2\times10^6$ M$_{\odot}$, twice that of Ret II. If the $r$-process elements were produced via the same mechanism in Ret II and Tuc III, then either Tuc III had twice the gas mass to pollute or half the amount of $r$-process material. Current data do not distinguish between these possibilities, but we note that the ejecta mass from a neutron star merger could well vary from event to event \citep{cote2018}.  It is also worth noting that the stellar mass of Ret II is $2.6 \pm 0.2 \times 10^{3} M_{\odot}$ \citep{bechtol_des}, that of Tuc III is $0.8 \pm 0.1 \times 10^{3} M_{\odot}$ \citep{adw}, and that of the Tuc III stream is $3.8 \times 10^{3} M_{\odot}$ \citep{shipp}, so many possible dilution scenarios are plausible.

%----------
%The M$_eu$ = $10^-4.5$, estimate in Ji 2016 comes from [Eu/H]= -1 and assuming a gas mass of $10^6$ solar masses, so either we have less Eu in Tuc III or we have more gas. Simon 2015 list a mass of $5.6x10^5$ solar masses for Ret II, is this mass all; dark matter, stars, gas Nora Shipp gives a stellar mass of $2.6x10^3$ solar masses and progenitor mass of $8x10^4$ solar masses for Tuc III, how do these compare to the mass of Ret II?
%--------

\cite{sakaria} show that $r$-process enhanced stars in the Milky Way halo at a range of metallicities have nearly identical $r$-process patterns, matching the solar system $r$-process pattern, regardless of the level of $r$-process enhancement. \cite{hansen} also found this to be true for the $r$-process enhanced stars detected in classical and ultra faint dwarf galaxies. The implication of this result is that there must either be a single mechanism for $r$-process element production, or else every $r$-process production mechanism must produce identical abundance patterns. If the former, the most likely site of $r$-process enhancement is binary neutron star mergers, based on observational evidence to date.  Furthermore, in galaxies with masses as low as the ultra-faint dwarfs discussed here, the star formation history must have proceeded in such a way that there was likely only one enrichment event, early in the history of the galaxy \citep[e.g.][]{ojima}. The $r$-process enhancement of Tuc III is consistent with this result, although with a different $r$-process element or dilution gas mass than in the case of Ret II.

%\cite{sakaria} show that $r$-process enhanced stars at a range of metallicities have nearly identical $r$-process patterns, regardless of the level of $r$-process enhancement.  The implication of this result is that there must either be a single mechanism for $r$-process element production, or else every $r$-process production mechanism must produce identical abundance patterns.  If the former, the most likely site of $r$-process enhancement is binary neutron star mergers, based on observational evidence to date.  Furthermore, in galaxies with masses as low as the ultra-faint dwarfs discussed here, the star formation history must have proceeded in such a way that there was likely only one enrichment event, early in the history of the galaxy.  The $r$-process enhancement of Tuc III is consistent with this result, although the $r$-process elements must have been more diluted than in the case of Ret II.

\cite{hansen} noted that with a sample of one star it is difficult to determine whether the source of enhancement in this galaxy is inside or outside the galaxy.  With this larger sample of stars in both the core and the tidal tails, we can now claim that the abundance pattern of Tuc III, like that of Ret II, must be due to an enhancement event inside the galaxy.  Since it appears that stars throughout the galaxy show similar levels of $r$-process enhancement, we can further state that this enhancement event must have occurred early in the history of the galaxy, thereby polluting the galaxy on large scales.

\cite{scannapieco} used an adaptive mesh refinement cosmological simulation to show that the exact location of an enrichment event in small ultra-faint dwarf galaxies such as Ret II can have a large impact on the enrichment of all stars in the galaxy.  In the case of Ret II, they compared the results of a binary neutron star merger located at the center of the galaxy compared to on the outskirts of the galaxy to show that the high levels of $r$-process enhancement seen in Ret II can only be explained if the event occurred very close to the center of the galaxy, and at a time at which the stars were still being formed.  Another implication of the work of \cite{scannapieco} is that a galaxy with lower levels of $r$-process enhancement, such as Tuc III, may have experienced a similar nucleosynthetic event, but that it was located at the edges of the galaxy.  Such events occurring on the edges of the galaxy are not particularly unexpected, given the ``kicks'' binary neutron stars experience.  Now that multiple stars in Tuc III have been shown to be moderately enhanced in $r$-process elements, a more detailed comparison can be made to this theoretical work.

\subsection{Galaxy or Globular Cluster?}

Both \cite{simon} and \cite{ting} considered in some detail the nature of Tuc III, since the low measured velocity dispersion ($0.1^{+0.7}_{-0.1}$ km s$^{-1}$) leads to speculation as to whether Tuc III was truly an ultra-faint dwarf galaxy or rather a globular cluster at birth.
A possible reason for the low measured velocity dispersion, as discussed by \cite{simon}, is that stripping of the stars has lowered the velocity dispersion as Tuc III merges with the Galaxy. Since the velocity dispersion may not clearly determine the nature of Tuc III, we consider here several chemical aspects of the stellar population that may shed light on its origin.

\begin{figure}[htb!]
\centering 
\includegraphics[scale=0.48]{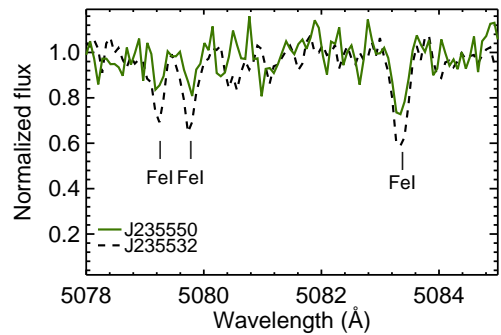}
\caption{Comparison of the spectrum studied by \cite{hansen} (DES J235532) and of one of the stars in this work, DES J235550. The temperatures and surface gravities of these two stars are very similar; the metallicity difference between the two stars is apparent.
}
\label{fig:com}
\end{figure}

The four stars studied here have a mean metallicity of [Fe/H]$\sim -$2.64 $\pm$ 0.15, significantly lower than the star studied by \cite{hansen} ([Fe/H]$\sim -$2.25).  Figure \ref{fig:com} compares two of these spectra and demonstrates that Tuc III is not a monometallic system.
%Our metallicities are consistent with those derived using the Calcium triplet in moderate resolution spectroscopy by \cite{simon} and \cite{ting} as shown in Figure \ref{fig:feh}.
Confirmation of Tuc III's metallicity dispersion using the five stars studied here provides further evidence that Tuc III is in fact a galaxy, since only galaxies, and not globular clusters, have gravitational wells deep enough to retain supernova ejecta, enabling the production of multiple generations of stars.  The resulting range of metallicities observed in galaxies is the natural result of this extended formation \citep{tinsley, willman}.  
We note that if we were to use the photometric temperature for DES J235532 we would derive a $\sim0.2$dex lower metallicity for this star.  Consequently, using the photometric temperatures to consider the metallicities of these stars would result in no statistically significant metallicity range between the stars, weakening the evidence that Tuc III is a galaxy and not a globular cluster.

\begin{figure}
\centering 
\includegraphics[scale=0.6]{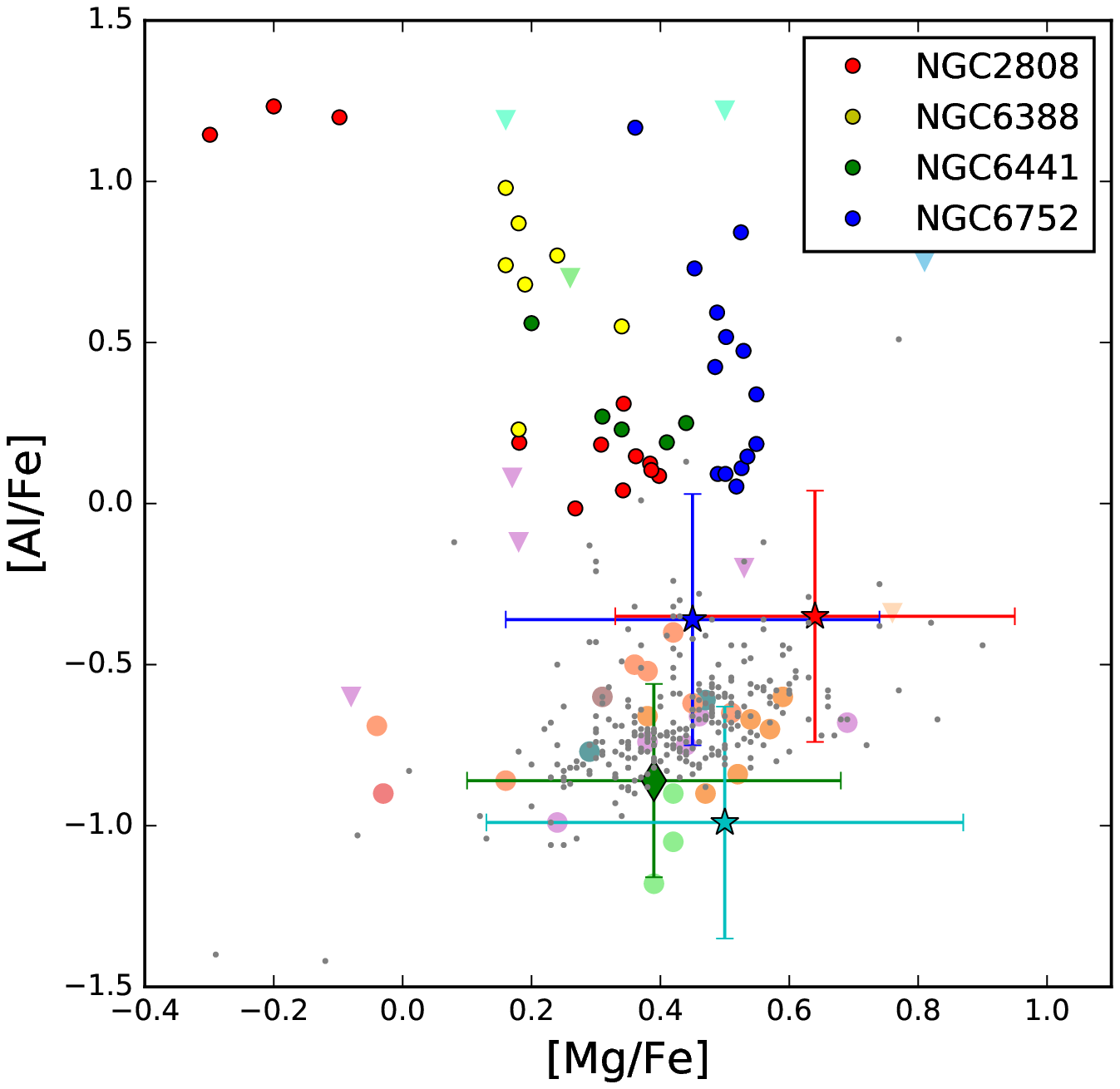}
\caption{
[Al/Fe] as a function of [Mg/Fe] for Tuc III stars compared to red giant stars in globular clusters NGC 2808 \citep{carretta}, NGC 6388 \citep{carretta_ngc6388}, NGC 6441\citep{gratton}, and NGC6752 \citep{carretta}, other ultra-faint dwarf galaxies (symbols as in previous figures), and stars in the halo \citep[grey points;]{roed}.
Tuc III does not appear to have a Mg-Al anti-correlation, as would be expected if it were a globular cluster. 
}
\label{fig:mgal}
\end{figure}

We further investigate the nature of Tuc III by considering the abundances of elements involved in proton-capture reactions, specifically Mg and Al.  The Mg--Al anticorrelation that is observed in globular clusters is thought to be produced via pollution of second generation stars in the cluster by massive asymptotic giant branch (AGB) stars at the end of their lives, particularly in massive or very metal-poor clusters \citep[e.g.][]{carretta}.  Figure \ref{fig:mgal} compares the Mg and Al abundances of the Tuc III stars to red giant stars in four globular clusters that have been shown to have a strong Mg--Al anticorrelation.  The Tuc III stars studied here do not exhibit the proton burning trend.

Finally, while it is true that globular clusters generally have neutron-capture enhancement similar to that observed here \citep[and very different from other ultra-faint dwarf galaxies, see][for example]{ji_grui}, we do not feel that this commonality is enough to claim that Tuc III is more likely to be a globular cluster. Furthermore, Tuc III's average metallicity, [Fe/H]$\sim$ $-$2.49 \citep{ting}, would place it on the extreme low end of the distribution of globular cluster metallicities: the lowest metallicity globular cluster in the \cite{harris, harris2010} catalog is NGC 7078 with [Fe/H]=$-$2.37.
We therefore conclude that Tuc III is most likely an ultra-faint dwarf galaxy and not a globular cluster.

%\subsection{Neutron-capture Element Enhancement}

%that, as summarized by \cite{hansen}, multiple classical dwarf galaxies in and around the Milky Way have been shown to contain single $r$-process enhanced stars. Furthermore, by now there are many Milky Way halo stars that show $r$-process enhanced abundances (ref Beers, Sneden).

%Notes from Terese (very old):
%Where did the extra Barium come from?
%There has been some r-process event in the past. Assume all the Eu was produced in that.    scale that to your abundance pattern.  then subtract that from the other elements. That should tell you what the other event could have been.
%Sr is s-process, first peak.  Yields from s-process are more certain.
%Reference: Roederer does this.

%\subsection{Location of enhancement event}

\subsection{Binarity}
\label{sec:binary}
Since all five of our stars were studied using multiple observations over a span of two years, we can use multi-epoch radial velocities to search for reflex motion due to the stars being in an undetected binary system. In Figure~\ref{fig:rv} we compare the radial velocities measured in this work with those measured by \cite{simon} and \cite{ting}.  We see very good agreement between the measurements of three of the stars: DES J235532, DES J235550, and DES J000549 do not appear have variable radial velocities according to these measurements and are therefore unlikely to be binaries with periods $\la$ 1 year.  Two other stars may have unseen binary companions: DES J234351 shows large radial velocity variation, and DES J235738 may also show small velocity variation. 

DES J234351 is identified as a binary by \cite{ting} as well, who note a $\Delta$v$\sim$20 km s$^{-1}$ and exclude the star from further analysis of Tuc III's kinematics.  Our higher precision velocities confirm the velocity shift between observations made in 2016 and 2017 and indicate that DES J234351 is very likely a binary system.

DES J235738 shows potential velocity variation with amplitude of  $\geq$2.79 km s$^{-1}$.  
%A simple chi-squared calculation shows that the deviation of our measured velocities from a constant value has a chi-squared value of 7.09 with 5 degrees of freedom, or a reduced chi-squared value of 1.42, where a value of 1 is the expectation for nonvariable velocity.  This corresponds to a p-value of 0.21, and the null hypothesis is rejected at the 79\% confidence interval.  
This (weakly) suggests that this star may in fact be a binary and warrants further kinematic measurements. If DES J235738 is shown to be in a binary system, mass transfer from its companion could potentially explain its chemistry as well, since this star appears to have some $s$-process enhancement in addition to the $r$-process enhancement shared with the other stars, although a higher S/N spectrum is needed to confirm this suggestion.

These are not the first binary stars discovered in an ultra-faint Milky Way satellite galaxy.  \cite{koch_hansen} measured velocities of one star in the Hercules dwarf galaxy over a two year baseline and concluded that it was in fact a binary system having a 135 day period, composed of a giant with a low-mass companion, likely a white dwarf.  Despite the fact that mass transfer binaries can explain peculiar abundance patterns in some cases, no such binary scenario could be described by \cite{koch_hansen} in the case of Hercules.  Conversely, a binary star in Segue 1 does show signs of mass transfer \citep{segue1} through its high carbon abundance.  Binary stars have also been detected via variable radial velocity signatures in Boo II \citep{booii}, Tri II \citep{venn, kirby}, Carina II \citep{li_carii}, and Ret II \citep{minor, simon_ret2}.

%\subsection{Barium} Where did the extra barium come from?

\section{Conclusions}
\label{section:concl}

We have presented chemical abundance measurements of four additional confirmed member stars in the  Tuc III stellar system: two stars located in the core of the galaxy and two in the tidal tails. Together with the star studied by \cite{hansen}, the sample of five stars shows that Tuc III is moderately enhanced in $r$-process elements ($r$-I), shows the expected trend in $\alpha$-elements, and is not carbon enhanced.  At least one, and possibly two, of the stars are likely to be binaries.  The abundance patterns of these stars suggest that Tuc III is an ultra-faint dwarf galaxy and not a globular cluster.

As can be seen in Figure~\ref{fig:cmd}, there are more than ten additional confirmed member stars of Tuc III that are bright enough to be studied in this way with today's largest telescopes and could be added to this sample: three blue horizontal branch stars and eight additional stars on the giant branch with g$<$19 \citep[although not all of these may be true members, see][]{pace}.  In the near term, study of these stars could increase the sample somewhat, until the next generation of telescopes enables the study of additional, fainter stars at high resolution.

Interestingly, two of the recently discovered southern hemisphere ultra-faint dwarf galaxies, Ret II and Tuc III, have now been shown to have multiple stars enhanced in $r$-process elements to a greater or lesser extent.  The other DES-discovered ultra-faint dwarfs that have been studied chemically to date, Tuc II \citep{ji_tucii}, Gru I \citep{ji_grui}, and Hor I \citep{nagasawa}, do not show $r$-process enhancement.  Additional ultra-faint dwarfs have member stars that are bright enough to be studied chemically, and may further add to the census of $r$-process enhanced galaxies.  The reason that Ret II and Tuc III, and none of the other galaxies, have multiple $r$-process enhanced stars is as yet unknown, but may become clearer with study of additional stars in these and other galaxies.

\acknowledgements{

The authors thank the referee for a careful reading of the manuscript and are grateful to A. McWilliam for insightful comments.  T. T. Hansen and J. D. Simon acknowledge support from National Science Foundation grant AST-1714873.

Funding for the DES Projects has been provided by the U.S. Department of Energy, the U.S. National Science Foundation, the Ministry of Science and Education of Spain, 
the Science and Technology Facilities Council of the United Kingdom, the Higher Education Funding Council for England, the National Center for Supercomputing 
Applications at the University of Illinois at Urbana-Champaign, the Kavli Institute of Cosmological Physics at the University of Chicago, 
the Center for Cosmology and Astro-Particle Physics at the Ohio State University,
the Mitchell Institute for Fundamental Physics and Astronomy at Texas A\&M University, Financiadora de Estudos e Projetos, 
Funda{\c c}{\~a}o Carlos Chagas Filho de Amparo {\`a} Pesquisa do Estado do Rio de Janeiro, Conselho Nacional de Desenvolvimento Cient{\'i}fico e Tecnol{\'o}gico and 
the Minist{\'e}rio da Ci{\^e}ncia, Tecnologia e Inova{\c c}{\~a}o, the Deutsche Forschungsgemeinschaft and the Collaborating Institutions in the Dark Energy Survey. 

The Collaborating Institutions are Argonne National Laboratory, the University of California at Santa Cruz, the University of Cambridge, Centro de Investigaciones Energ{\'e}ticas, 
Medioambientales y Tecnol{\'o}gicas-Madrid, the University of Chicago, University College London, the DES-Brazil Consortium, the University of Edinburgh, 
the Eidgen{\"o}ssische Technische Hochschule (ETH) Z{\"u}rich, 
Fermi National Accelerator Laboratory, the University of Illinois at Urbana-Champaign, the Institut de Ci{\`e}ncies de l'Espai (IEEC/CSIC), 
the Institut de F{\'i}sica d'Altes Energies, Lawrence Berkeley National Laboratory, the Ludwig-Maximilians Universit{\"a}t M{\"u}nchen and the associated Excellence Cluster Universe, 
the University of Michigan, the National Optical Astronomy Observatory, the University of Nottingham, The Ohio State University, the University of Pennsylvania, the University of Portsmouth, 
SLAC National Accelerator Laboratory, Stanford University, the University of Sussex, Texas A\&M University, and the OzDES Membership Consortium.

Based in part on observations at Cerro Tololo Inter-American Observatory, National Optical Astronomy Observatory, which is operated by the Association of 
Universities for Research in Astronomy (AURA) under a cooperative agreement with the National Science Foundation.

The DES data management system is supported by the National Science Foundation under Grant Numbers AST-1138766 and AST-1536171.
The DES participants from Spanish institutions are partially supported by MINECO under grants AYA2015-71825, ESP2015-66861, FPA2015-68048, SEV-2016-0588, SEV-2016-0597, and MDM-2015-0509, 
some of which include ERDF funds from the European Union. IFAE is partially funded by the CERCA program of the Generalitat de Catalunya.
Research leading to these results has received funding from the European Research
Council under the European Union's Seventh Framework Program (FP7/2007-2013) including ERC grant agreements 240672, 291329, and 306478.
We  acknowledge support from the Australian Research Council Centre of Excellence for All-sky Astrophysics (CAASTRO), through project number CE110001020, and the Brazilian Instituto Nacional de Ci\^encia
e Tecnologia (INCT) e-Universe (CNPq grant 465376/2014-2).

This manuscript has been authored by Fermi Research Alliance, LLC under Contract No. DE-AC02-07CH11359 with the U.S. Department of Energy, Office of Science, Office of High Energy Physics. The United States Government retains and the publisher, by accepting the article for publication, acknowledges that the United States Government retains a non-exclusive, paid-up, irrevocable, world-wide license to publish or reproduce the published form of this manuscript, or allow others to do so, for United States Government purposes.
}

\input{abundtable.tab}
\input{errortable.tab}

\input{rv.tab}

\end{document}

%% file: authors.tex
\author{J. L. Marshall}
\affil{Mitchell Institute for Fundamental Physics and Astronomy and Department of Physics and Astronomy, Texas A\&M University, College Station, TX 77843-4242, USA\\}

\author{T. Hansen}
\affil{Carnegie Institution for Science, 813 Santa Barbara St., Pasadena, CA 91101, USA\\}

\author{J. D. Simon}
\affil{Carnegie Institution for Science, 813 Santa Barbara St., Pasadena, CA 91101, USA\\}

\author{T. S. Li}
\affil{Fermi National Accelerator Laboratory, P. O. Box 500, Batavia, IL 60510, USA\\}

\author{R. A. Bernstein} 
\affil{Carnegie Institution for Science, 813 Santa Barbara St., Pasadena, CA 91101, USA\\}

\author{K. Kuehn}
\affil{Australian Astronomical Optics, Macquarie University, North Ryde, NSW 2113, Australia}

\author{A. B. Pace}
\affil{Mitchell Institute for Fundamental Physics and Astronomy and Department of Physics and Astronomy, Texas A\&M University, College Station, TX 77843-4242, USA\\}

\author{D. L. DePoy}
\affil{Mitchell Institute for Fundamental Physics and Astronomy and Department of Physics and Astronomy, Texas A\&M University, College Station, TX 77843-4242, USA\\}

\author{A. Palmese}
\affil{Fermi National Accelerator Laboratory, P. O. Box 500, Batavia, IL 60510, USA\\}

\author{A. Pieres}
\affil{Laborat\'orio Interinstitucional de e-Astronomia - LIneA, Rua Gal. Jos\'e Cristino 77, Rio de Janeiro, RJ - 20921-400, Brazil\\}

\author{L. Strigari}
\affil{Mitchell Institute for Fundamental Physics and Astronomy and Department of Physics and Astronomy, Texas A\&M University, College Station, TX 77843-4242, USA\\}

\author{A. Drlica-Wagner}
\affil{Fermi National Accelerator Laboratory, P. O. Box 500, Batavia, IL 60510, USA\\}

\author{K. Bechtol}
\affil{LSST, 933 North Cherry Avenue, Tucson, AZ 85721, USA\\}
\affil{Physics Department, 2320 Chamberlin Hall, University of Wisconsin-Madison, 1150 University Avenue Madison, WI  53706-1390\\}

\author{C.~Lidman}
\affil{The Research School of Astronomy and Astrophysics, Australian National University, ACT 2601, Australia\\}

\author{D. Q. Nagasawa}
\affil{Mitchell Institute for Fundamental Physics and Astronomy and Department of Physics and Astronomy, Texas A\&M University, College Station, TX 77843-4242, USA\\}

\author{E.~Bertin}
\affil{CNRS, UMR 7095, Institut d'Astrophysique de Paris, F-75014, Paris, France}
\affil{Sorbonne Universit\'es, UPMC Univ Paris 06, UMR 7095, Institut d'Astrophysique de Paris, F-75014, Paris, France}
\author{D.~Brooks}
\affil{Department of Physics \& Astronomy, University College London, Gower Street, London, WC1E 6BT, UK}
\author{E.~Buckley-Geer}
\affil{Fermi National Accelerator Laboratory, P. O. Box 500, Batavia, IL 60510, USA}
\author{D.~L.~Burke}
\affil{Kavli Institute for Particle Astrophysics \& Cosmology, P. O. Box 2450, Stanford University, Stanford, CA 94305, USA}
\affil{SLAC National Accelerator Laboratory, Menlo Park, CA 94025, USA}
\author{A.~Carnero~Rosell}
\affil{Centro de Investigaciones Energ\'eticas, Medioambientales y Tecnol\'ogicas (CIEMAT), Madrid, Spain}
\affil{Laborat\'orio Interinstitucional de e-Astronomia - LIneA, Rua Gal. Jos\'e Cristino 77, Rio de Janeiro, RJ - 20921-400, Brazil}
\author{M.~Carrasco~Kind}
\affil{Department of Astronomy, University of Illinois at Urbana-Champaign, 1002 W. Green Street, Urbana, IL 61801, USA}
\affil{National Center for Supercomputing Applications, 1205 West Clark St., Urbana, IL 61801, USA}
\author{J.~Carretero}
\affil{Institut de F\'{\i}sica d'Altes Energies (IFAE), The Barcelona Institute of Science and Technology, Campus UAB, 08193 Bellaterra (Barcelona) Spain}
\author{C.~E.~Cunha}
\affil{Kavli Institute for Particle Astrophysics \& Cosmology, P. O. Box 2450, Stanford University, Stanford, CA 94305, USA}
\author{C.~B.~D'Andrea}
\affil{Department of Physics and Astronomy, University of Pennsylvania, Philadelphia, PA 19104, USA}
\author{L.~N.~da Costa}
\affil{Laborat\'orio Interinstitucional de e-Astronomia - LIneA, Rua Gal. Jos\'e Cristino 77, Rio de Janeiro, RJ - 20921-400, Brazil}
\affil{Observat\'orio Nacional, Rua Gal. Jos\'e Cristino 77, Rio de Janeiro, RJ - 20921-400, Brazil}
\author{J.~De~Vicente}
\affil{Centro de Investigaciones Energ\'eticas, Medioambientales y Tecnol\'ogicas (CIEMAT), Madrid, Spain}
\author{S.~Desai}
\affil{Department of Physics, IIT Hyderabad, Kandi, Telangana 502285, India}
\author{P.~Doel}
\affil{Department of Physics \& Astronomy, University College London, Gower Street, London, WC1E 6BT, UK}
\author{T.~F.~Eifler}
\affil{Department of Astronomy/Steward Observatory, 933 North Cherry Avenue, Tucson, AZ 85721-0065, USA}
\affil{Jet Propulsion Laboratory, California Institute of Technology, 4800 Oak Grove Dr., Pasadena, CA 91109, USA}
\author{B.~Flaugher}
\affil{Fermi National Accelerator Laboratory, P. O. Box 500, Batavia, IL 60510, USA}
\author{P.~Fosalba}
\affil{Institut d'Estudis Espacials de Catalunya (IEEC), 08034 Barcelona, Spain}
\affil{Institute of Space Sciences (ICE, CSIC),  Campus UAB, Carrer de Can Magrans, s/n,  08193 Barcelona, Spain}
\author{J.~Frieman}
\affil{Fermi National Accelerator Laboratory, P. O. Box 500, Batavia, IL 60510, USA}
\affil{Kavli Institute for Cosmological Physics, University of Chicago, Chicago, IL 60637, USA}
\author{J.~Garc\'ia-Bellido}
\affil{Instituto de Fisica Teorica UAM/CSIC, Universidad Autonoma de Madrid, 28049 Madrid, Spain}
\author{E.~Gaztanaga}
\affil{Institut d'Estudis Espacials de Catalunya (IEEC), 08034 Barcelona, Spain}
\affil{Institute of Space Sciences (ICE, CSIC),  Campus UAB, Carrer de Can Magrans, s/n,  08193 Barcelona, Spain}
\author{D.~W.~Gerdes}
\affil{Department of Astronomy, University of Michigan, Ann Arbor, MI 48109, USA}
\affil{Department of Physics, University of Michigan, Ann Arbor, MI 48109, USA}
\author{R.~A.~Gruendl}
\affil{Department of Astronomy, University of Illinois at Urbana-Champaign, 1002 W. Green Street, Urbana, IL 61801, USA}
\affil{National Center for Supercomputing Applications, 1205 West Clark St., Urbana, IL 61801, USA}
\author{J.~Gschwend}
\affil{Laborat\'orio Interinstitucional de e-Astronomia - LIneA, Rua Gal. Jos\'e Cristino 77, Rio de Janeiro, RJ - 20921-400, Brazil}
\affil{Observat\'orio Nacional, Rua Gal. Jos\'e Cristino 77, Rio de Janeiro, RJ - 20921-400, Brazil}
\author{G.~Gutierrez}
\affil{Fermi National Accelerator Laboratory, P. O. Box 500, Batavia, IL 60510, USA}
\author{W.~G.~Hartley}
\affil{Department of Physics \& Astronomy, University College London, Gower Street, London, WC1E 6BT, UK}
\affil{Department of Physics, ETH Zurich, Wolfgang-Pauli-Strasse 16, CH-8093 Zurich, Switzerland}
\author{D.~L.~Hollowood}
\affil{Santa Cruz Institute for Particle Physics, Santa Cruz, CA 95064, USA}
\author{K.~Honscheid}
\affil{Center for Cosmology and Astro-Particle Physics, The Ohio State University, Columbus, OH 43210, USA}
\affil{Department of Physics, The Ohio State University, Columbus, OH 43210, USA}
\author{B.~Hoyle}
\affil{Max Planck Institute for Extraterrestrial Physics, Giessenbachstrasse, 85748 Garching, Germany}
\affil{Universit\"ats-Sternwarte, Fakult\"at f\"ur Physik, Ludwig-Maximilians Universit\"at M\"unchen, Scheinerstr. 1, 81679 M\"unchen, Germany}
\author{D.~J.~James}
\affil{Harvard-Smithsonian Center for Astrophysics, Cambridge, MA 02138, USA}
\author{N.~Kuropatkin}
\affil{Fermi National Accelerator Laboratory, P. O. Box 500, Batavia, IL 60510, USA}

\author{M.~A.~G.~Maia}
\affil{Laborat\'orio Interinstitucional de e-Astronomia - LIneA, Rua Gal. Jos\'e Cristino 77, Rio de Janeiro, RJ - 20921-400, Brazil}
\affil{Observat\'orio Nacional, Rua Gal. Jos\'e Cristino 77, Rio de Janeiro, RJ - 20921-400, Brazil}
\author{F.~Menanteau}
\affil{Department of Astronomy, University of Illinois at Urbana-Champaign, 1002 W. Green Street, Urbana, IL 61801, USA}
\affil{National Center for Supercomputing Applications, 1205 West Clark St., Urbana, IL 61801, USA}
\author{C.~J.~Miller}
\affil{Department of Astronomy, University of Michigan, Ann Arbor, MI 48109, USA}
\affil{Department of Physics, University of Michigan, Ann Arbor, MI 48109, USA}
\author{R.~Miquel}
\affil{Instituci\'o Catalana de Recerca i Estudis Avan\c{c}ats, E-08010 Barcelona, Spain}
\affil{Institut de F\'{\i}sica d'Altes Energies (IFAE), The Barcelona Institute of Science and Technology, Campus UAB, 08193 Bellaterra (Barcelona) Spain}
\author{A.~A.~Plazas}
\affil{Jet Propulsion Laboratory, California Institute of Technology, 4800 Oak Grove Dr., Pasadena, CA 91109, USA}
\author{E.~Sanchez}
\affil{Centro de Investigaciones Energ\'eticas, Medioambientales y Tecnol\'ogicas (CIEMAT), Madrid, Spain}
\author{B.~Santiago}
\affil{Instituto de F\'\i sica, UFRGS, Caixa Postal 15051, Porto Alegre, RS - 91501-970, Brazil}
\affil{Laborat\'orio Interinstitucional de e-Astronomia - LIneA, Rua Gal. Jos\'e Cristino 77, Rio de Janeiro, RJ - 20921-400, Brazil}
\author{V.~Scarpine}
\affil{Fermi National Accelerator Laboratory, P. O. Box 500, Batavia, IL 60510, USA}
\author{M.~Schubnell}
\affil{Department of Physics, University of Michigan, Ann Arbor, MI 48109, USA}
\author{S.~Serrano}
\affil{Institut d'Estudis Espacials de Catalunya (IEEC), 08034 Barcelona, Spain}
\affil{Institute of Space Sciences (ICE, CSIC),  Campus UAB, Carrer de Can Magrans, s/n,  08193 Barcelona, Spain}
\author{I.~Sevilla-Noarbe}
\affil{Centro de Investigaciones Energ\'eticas, Medioambientales y Tecnol\'ogicas (CIEMAT), Madrid, Spain}
\author{M.~Smith}
\affil{School of Physics and Astronomy, University of Southampton,  Southampton, SO17 1BJ, UK}
\author{M.~Soares-Santos}
\affil{Brandeis University, Physics Department, 415 South Street, Waltham MA 02453}
\author{E.~Suchyta}
\affil{Computer Science and Mathematics Division, Oak Ridge National Laboratory, Oak Ridge, TN 37831}
\author{M.~E.~C.~Swanson}
\affil{National Center for Supercomputing Applications, 1205 West Clark St., Urbana, IL 61801, USA}
\author{G.~Tarle}
\affil{Department of Physics, University of Michigan, Ann Arbor, MI 48109, USA}
\author{W.~Wester}
\affil{Fermi National Accelerator Laboratory, P. O. Box 500, Batavia, IL 60510, USA}

\collaboration{(DES Collaboration)}

%% file: Fe.tab
\begin{deluxetable*}{llrrrrr}
\tablecaption{Fe~I and Fe~II lines used for parameter determination. \label{tab:ew}}
\tablewidth{0pt}
%\tabletypesize{\scriptsize}
\tablehead{\colhead{Stellar ID} & \colhead{Species} & \colhead{$\lambda$} & \colhead{$\chi$} & \colhead{$\log$ $gf$} & \colhead{EW} & \colhead{$\log \epsilon$}\\ 
\colhead{} & \colhead{} & \colhead{(\AA)} & \colhead{(eV)} & \colhead{} & \colhead{(m\AA)} & \colhead{}}
\startdata
J000549-593406 &\ion{Fe}{1} & 3765.54 &  3.240 &    0.480 &   78.63 &  4.638\\   
J000549-593406 &\ion{Fe}{1} & 3790.09 &  0.989 & $-$1.740 &  106.00 &  4.988\\   
J000549-593406 &\ion{Fe}{1} & 3805.34 &  3.300 &    0.310 &   79.81 &  4.913\\   
J000549-593406 &\ion{Fe}{1} & 3807.54 &  2.221 & $-$0.990 &   80.07 &  4.934\\   
J000549-593406 &\ion{Fe}{1} & 3876.04 &  1.010 & $-$2.890 &   74.29 &  5.158\\   
J000549-593406 &\ion{Fe}{1} & 3917.18 &  0.989 & $-$2.150 &  101.40 &  5.122\\   
J000549-593406 &\ion{Fe}{1} & 4067.98 &  3.209 & $-$0.530 &   41.48 &  4.630\\   
%\multicolumn{7}{c}{$\cdots$} \\
\multicolumn{1}{c}{\vdots} & \multicolumn{1}{l}{\phantom{0}\vdots} & \multicolumn{1}{c}{\vdots} & \multicolumn{1}{c}{\vdots} & \multicolumn{1}{c}{\vdots} & \multicolumn{1}{c}{\vdots} & \multicolumn{1}{c}{\vdots} \\
J000549-593406 &\ion{Fe}{2}&  4233.17 & 2.580 & $-$1.810 &   81.79 &   4.746\\
J000549-593406 &\ion{Fe}{2}&  4416.83 & 2.780 & $-$2.410 &   57.10 &   4.979\\
J000549-593406 &\ion{Fe}{2}&  4489.18 & 2.828 & $-$2.971 &   22.98 &   4.869\\
J000549-593406 &\ion{Fe}{2}&  4491.41 & 2.850 & $-$2.640 &   28.78 &   4.702\\
\enddata
\tablecomments{
The complete version of Table~\ref{tab:ew} is available online only.
A short version is shown here to illustrate its form and content.
}
\end{deluxetable*}

%% file: abundtable.tab
\begin{deluxetable*}{lcccccccccccccccc}
\tablecaption{Measured abundances\label{abund}} 
\tablehead{
&\multicolumn{4}{c}{J000549}&\multicolumn{4}{c}{J234351}&\multicolumn{4}{c}{J235550}&\multicolumn{4}{c}{J235738} \\
Element & [X/Fe] & log $\epsilon$(X) & $\sigma$ & N
& [X/Fe] & log $\epsilon$(X) & $\sigma$ & N
& [X/Fe] & log $\epsilon$(X) & $\sigma$ & N
& [X/Fe] & log $\epsilon$(X) & $\sigma$ & N \\
}
\startdata
C    & $-$0.43 &  5.39     &$\cdots$ &     &   $-$0.28 &      5.46 & $\cdots$&    & $-$0.19  &   5.55   &$\cdots$ &      & $-$0.05  &   5.80  &$\cdots$&     \\    
N    & $<$2.00 & $<$7.22   &$\cdots$ &     &  $<$ 1.5  &   $<$6.64 &$\cdots$ &    & $<$1.0   &  $<$6.14 &$\cdots$ &      &  $<$0.5  & $<$5.75 &$\cdots$&     \\    
NaI  & $\cdots$&$\cdots$   &$\cdots$ &     &      0.32 &      3.87 & 0.35    &  2 & $\cdots$ &$\cdots$  &$\cdots$ &      &    0.53  &   4.19  & 0.33   & 2   \\    
MgI  &    0.45 &      5.44 & 0.29    &  9  &      0.50 &      5.41 & 0.37    &  7 &     0.40 &    5.31  & 0.27    &  8   &    0.64  &   5.66  & 0.31   & 5   \\    
AlI  & $-$0.36 &      3.48 & 0.39    &  2  &   $-$0.99 &      2.78 & 0.36    &  2 & $\cdots$ &$\cdots$  &$\cdots$ &  2   & $-$0.35  &   3.52  &$\cdots$& 1   \\    
SiI  &    0.39 &      5.29 & 0.33    &  2  &      0.18 &      5.00 & 0.33    &  2 &     0.39 &    5.21  & 0.31    &  2   &    0.00  &   4.93  &$\cdots$& 1   \\    
KI   &    0.71 &      3.13 & 0.25    &  2  &      0.41 &      2.75 & $\cdots$&  1 &     0.71 &    3.05  &$\cdots$ &  1   &    0.56  &   3.01  & 0.31   & 2   \\    
CaI  &    0.40 &      4.13 & 0.28    &  19 &      0.35 &      4.00 & 0.28    &  11&     0.44 &    4.09  & 0.28    &  13  &    0.46  &   4.22  & 0.28   & 13  \\    
ScII &    0.15 &      0.69 & 0.28    &  5  &      0.09 &      0.55 & 0.27    &  4 &  $-$0.01 &    0.45  & 0.28    &  6   &    0.05  &   0.62  & 0.30   & 6   \\    
TiI  &    0.07 &      2.41 & 0.31    &  18 &      0.17 &      2.43 & 0.34    &  7 &     0.15 &    2.41  & 0.36    &  12  &    0.28  &   2.65  & 0.32   & 8   \\    
TiII &    0.46 &      2.80 & 0.29    &  39 &      0.31 &      2.57 & 0.29    &  11&     0.50 &    2.76  & 0.31    &  32  &    0.41  &   2.78  & 0.32   & 26  \\    
VI   & $-$0.16 &      1.16 &$\cdots$ &  1  &      0.06 &      1.30 &$\cdots$ &  1 &  $-$0.18 &    6.44  &$\cdots$ &  1   & $-$0.13  &   1.22  &$\cdots$& 1   \\    
VII  &    0.07 &      1.39 & 0.23    &  2  &      0.13 &      1.37 & 0.21    &  2 &     0.10 &    6.72  & 0.25    &  2   &    0.00  &   1.35  &$\cdots$& 1   \\    
CrI  & $-$0.38 &      2.65 & 0.32    &  14 &   $-$0.34 &      2.61 & 0.33    &  7 &  $-$0.26 &    2.69  & 0.34    &  9   & $-$0.17  &   2.89  & 0.34   & 6   \\    
CrII &    0.08 &      3.11 & 0.23    &  2  &      0.29 &      3.24 & 0.30    &  2 &     0.35 &    3.30  & 0.20    &  2   &    0.24  &   3.30  &$\cdots$& 1   \\    
MnI  & $-$0.50 &      2.32 & 0.24    &  6  &   $-$0.42 &      2.32 & 0.25    &  5 &  $-$0.50 &    2.25  & 0.23    &  6   & $-$0.39  &   2.46  & 0.23   & 4   \\    
FeI  & $-$2.61 &      4.89 & 0.17    &  157&   $-$2.69 &      4.81 & 0.20    &  90&  $-$2.69 &    4.81  & 0.21    &  122 & $-$2.58  &   4.92  & 0.20   & 103 \\    
FeII & $-$2.62 &      4.88 & 0.18    &  20 &   $-$2.70 &      4.80 & 0.17    &  16&  $-$2.69 &    4.81  & 0.08    &  13  & $-$2.59  &   4.91  & 0.14   & 15  \\    
CoI  &    0.03 &      2.41 & 0.30    &  3  &   0.14    &      2.44 & 0.29    &  3 &  $-$0.04 &    2.27  & 0.29    &  2   & $\cdots$ & $\cdots$&$\cdots$&     \\    
NiI  &    0.00 &      3.61 & 0.31    &  11 &   0.13    &      3.66 & 0.32    &  5 &     0.22 &    3.75  & 0.43    &  5   &    0.04  &   3.68  & 0.33   & 6   \\    
CuI  &$<-$0.5&    $<$1.28  &$\cdots$ &  1  &   $<$0.5  &    $<$2.00&$\cdots$ &  1 &  $<$0.0  & $<$1.50  &$\cdots$ &  1   & $-$0.20  &   1.41  &$\cdots$& 1   \\    
ZnI  & $-$0.08 &      1.87 &$\cdots$ &  1  &   0.33    &      2.20 & 0.18    &  2 &     0.19 &    2.06  &$\cdots$ &  1   &    0.22  &   2.20  & 0.22   & 2   \\    
SrII &   0.09  &      0.35 & 0.33    &  2  &  $-$0.26  &   $-$0.08 & 0.32    &  2 &     0.04 &    0.22  & 0.31    &  2   & $-$0.03  &   0.27  & 0.31   & 2   \\    
YII  &$-$0.20  &   $-$0.60 & 0.26    &  5  & $<-$0.3   &  $<-$0.78 &$\cdots$ &    &  $-$0.28 & $-$0.76  & 0.22    &  3   & $-$0.05  & $-$0.42 & 0.25   & 2   \\    
ZrII &    0.05 &      0.02 & 0.26    &  3  &      0.00 &   $-$0.11 &$\cdots$ &  1 &  $-$0.04 & $-$0.15  & 0.27    &  2   &    0.29  &   0.29  & 0.26   & 2   \\    
BaII &    0.11 &   $-$0.32 & 0.32    &  4  &   $-$0.20 &   $-$0.71 & 0.38    &  4 &  $-$0.05 & $-$0.56  & 0.24    &  4   &    0.14  & $-$0.26 & 0.34   & 4   \\    
LaII &    0.21 &   $-$1.30 & 0.23    &  2  &      0.25 &   $-$1.34 &$\cdots$ &  1 &     0.13 & $-$1.46  & 0.23    &  2   &    0.16  & $-$1.32 & 0.26   & 2   \\    
CeII &    0.32 &   $-$0.71 & 0.29    &  4  &   $<$1.0  &  $<-$0.11 &$\cdots$ &    &  $<$0.30 & $<-$0.81 & $\cdots$&      &  $<$0.3  &$<-$0.70 &$\cdots$&     \\    
PrII &    0.61 &   $-$1.28 &$\cdots$ &     &   $<$1.0  &  $<-$0.97 &$\cdots$ &    & $<$0.70  & $<-$1.27 &$\cdots$ &      &    0.50  & $-$1.36 & 0.23   & 3   \\    
NdII &    0.27 &   $-$0.92 & 0.24    &  8  &      0.69 &   $-$0.58 & 0.23    &  3 &  0.39    & $-$0.88  & 0.33    &  4   &    0.33  & $-$0.83 & 0.29   & 6   \\    
SmII &$<$0.50  &$<-$1.15   &$\cdots$ &     &   $<$0.8  &  $<-$0.93 &$\cdots$ &    & $<$0.70  & $<-$1.03 & $\cdots$&      &  $<$0.5  &$<-$1.12 &$\cdots$&     \\    
EuII &    0.49 &   $-$1.61 & 0.21    &  2  &      0.44 &   $-$1.73 & 0.29    &  2 &    0.56  & $-$1.61  & 0.22    &  3   &    0.22  & $-$1.84 & 0.21   & 3   \\    
GdII &$<$0.70  &  $<-$0.84 &$\cdots$ &     &   $<$1.3  &  $<-$0.32 &$\cdots$ &    & $<$1.00  &$<-$0.62  &$\cdots$ &      & $<$0.8   &$<-$0.71 &$\cdots$&     \\    
DyII &    0.59 &   $-$0.92 & 0.24    &  3  &   $<$0.7  &  $<-$0.89 &$\cdots$ &    & $<$1.00  &$<-$0.59  &$\cdots$ &      &    0.70  & $-$0.78 & 0.29   & 3   \\    
ErII &    0.50 &   $-$1.19 &$\cdots$ &  1  &   $<$1.0  &  $<-$0.77 &$\cdots$ &    & $<$0.60  &$<-$1.17  &$\cdots$ &      &  $\cdots$& $\cdots$&$\cdots$&     \\    
\enddata
\end{deluxetable*} 

%% file: errortable.tab
\begin{deluxetable}{lccccc}
\tablecaption{Uncertainties from stellar parameters for DES J00549\label{error}} 
\tablehead{
Element   &   $\Delta$T$_{eff}$   &   $\Delta$log $g$   &   $\Delta$V$_{mic}$   &   $\Delta$[Fe/H]   &   Total\\
}
\startdata
C   &   0.16   &   0.02   &   0.1   &   0.17   &   0.25 \\
Na   &   0.24   &   0.12   &   0.1   &   0.17   &   0.33 \\
MgI   &   0.14   &   0.05   &   0.05   &   0.17   &   0.23 \\
AlI   &   0.27   &   0.17   &   0.04   &   0.17   &   0.36 \\
SiI   &   0.23   &   0.03   &   0.12   &   0.17   &   0.31 \\
KI   &   0.16   &   0.03   &   0.07   &   0.17   &   0.25 \\
CaI   &   0.12   &   0.03   &   0.07   &   0.17   &   0.22 \\
ScI   &   0.07   &   0.08   &   0.18   &   0.17   &   0.27 \\
TiI   &   0.22   &   0.04   &   0.04   &   0.17   &   0.28 \\
TiII   &   0.06   &   0.08   &   0.13   &   0.17   &   0.24 \\
VI   &   0.21   &   0   &   0.02   &   0.17   &   0.27 \\
VII   &   0.07   &   0.1   &   0   &   0.17   &   0.21 \\
CrI   &   0.2   &   0.04   &   0.08   &   0.17   &   0.28 \\
CrII   &   0.03   &   0.1   &   0.02   &   0.17   &   0.2 \\
MnI   &   0.15   &   0.02   &   0.05   &   0.17   &   0.23 \\
CoI   &   0.2   &   0.1   &   0.05   &   0.17   &   0.29 \\
NiI   &   0.17   &   0.03   &   0.07   &   0.17   &   0.25 \\
ZnI   &   0.06   &   0.04   &   0   &   0.17   &   0.18 \\
Sr   &   0.15   &   0.05   &   0.2   &   0.17   &   0.31 \\
Y   &   0.06   &   0.1   &   0.02   &   0.17   &   0.21 \\
Zr   &   0.2   &   0   &   0.05   &   0.17   &   0.26 \\
Ba   &   0.09   &   0.07   &   0.01   &   0.17   &   0.2 \\
La   &   0.12   &   0.1   &   0   &   0.17   &   0.23 \\
Ce   &   0.1   &   0.1   &   0.01   &   0.17   &   0.22 \\
Pr   &   0.09   &   0.05   &   0.06   &   0.17   &   0.21 \\
Nd   &   0.11   &   0.09   &   0.02   &   0.17   &   0.22 \\
Eu   &   0.07   &   0.08   &   0.07   &   0.17   &   0.21 \\
Dy   &   0.13   &   0.09   &   0   &   0.17   &   0.23 \\
Er   &   0.11   &   0.13   &   0.07   &   0.17   &   0.25 \\
\enddata
\end{deluxetable}

%% file: rv.tab
\begin{deluxetable*}{llccc}
\tablecaption{Radial velocities from all sources\label{rvtable}} 
\tablehead{Object Name & Date Observed\tablenotemark{a} & v$_{hel}$& error & Reference}
\startdata
DES J235532 &  57248 & $-$103.4 & 0.3 & \cite{hansen}\\
&57223.3&$-$102.32&1.23&\cite{simon}\\
&57632.8&$-$103.26&1.00&\cite{simon}\\
&57589\tablenotemark{b}  &$-$102.89&0.51&\cite{ting}\\
\hline
DES J235738 &  57605.69 &  $-$100.9 & 0.8&this work\\
&57220.8&$-$102.24&1.21&\cite{simon}\\
&57223.3 &$-$100.3&1.27&\cite{simon}\\
&57312.2&$-$102.37&1.22&\cite{simon}\\
&57630.9&$-$101.76&1.00&\cite{simon}\\
&57589\tablenotemark{b}  & $-$99.58&0.71&\cite{ting}\\
\hline
DES J235550 &  57607.77& $-$102.9&0.8&this work\\
&57220.8&$-$102.88&1.21&\cite{simon}\\
&57223.3&$-$101.81&2.90&\cite{simon}\\
&57630.9&$-$101.92&1.00&\cite{simon}\\
&57589\tablenotemark{b}  &$-$101.94&0.40&\cite{ting}\\
\hline
DES J000549 & 57605.78 & $-$92.6&0.9& this work \\
&57589\tablenotemark{b}&$-$92.01&0.70&\cite{ting}\\
\hline
DES J234351 & 57606.82 &  $-$121.5&1.1&this work\\
&57589&$-$122.2&0.8&\cite{ting}\\
&57987&$-$99.4&1.4&\cite{ting}\\
\enddata
\tablenotetext{a}{Reported at the midpoint of the observation.}
\tablenotetext{b}{Average value over a ten night observing run.}
\end{deluxetable*}